\documentclass{article}

\PassOptionsToPackage{numbers,sort&compress}{natbib}

\usepackage[preprint]{neurips_2026}

\usepackage[utf8]{inputenc} %
\usepackage[T1]{fontenc}    %
\usepackage{hyperref}       %
\usepackage{url}            %
\usepackage{booktabs}       %
\usepackage{amsfonts}       %
\usepackage{nicefrac}       %
\usepackage{microtype}      %
\usepackage{xcolor}         %

\definecolor{citecolor}{RGB}{31,119,180} %
\hypersetup{colorlinks=true,citecolor=citecolor}

\usepackage{xspace}
\usepackage{longtable}
\usepackage{graphicx}
\usepackage{colortbl}
\usepackage{array}
\usepackage{tabularx}
\usepackage{multirow}
\usepackage{caption}
\usepackage{subcaption}
\usepackage{amssymb,amsmath}
\usepackage{hhline}
\usepackage{ragged2e}

\usepackage{mdframed}
\definecolor{blockquote-bg}{gray}{0.95}   %
\definecolor{blockquote-bar}{gray}{0.70}  %

\usepackage[capitalize,noabbrev]{cleveref}
\usepackage{appendix}
\usepackage{diagbox}
\usepackage{etoc}
\etocdepthtag.toc{mtchapter}
\etocsettagdepth{mtchapter}{subsection}
\etocsettagdepth{mtappendix}{none}

\usepackage{tikz}
\usepackage{wrapfig}
\usepackage{placeins}
\usepackage{needspace}
\usetikzlibrary{patterns}

\newmdenv[
  backgroundcolor=blockquote-bg,
  linecolor=blockquote-bar,
  leftline=true,
  rightline=false,
  topline=false,
  bottomline=false,
  linewidth=3pt,
  innerleftmargin=10pt,
  innerrightmargin=10pt,
  innertopmargin=6pt,
  innerbottommargin=6pt,
  roundcorner=2pt,        %
  skipabove=\baselineskip,
  skipbelow=\baselineskip
]{blockquote}

\newcommand{\eg}{\textit{e.g.}\xspace}

\newcommand{\etal}{\textit{et al.}\xspace}
\newcommand{\vct}[1]{\boldsymbol{#1}} 
 
\newlength\savewidth

\newcommand{\methodname}{MedVTok\xspace}

\renewcommand{\paragraph}[1]{\noindent\textbf{#1}\xspace}

\title{\methodname: A General-Purpose Medical Visual Tokenizer}

\newcommand{\paperauthors}{%
  \parbox[t]{\dimexpr\textwidth-2\tabcolsep\relax}{%
    \centering
    \normalsize\normalfont\bfseries
    \addtolength{\baselineskip}{1pt}
    \mbox{Chenglong Ma\textsuperscript{1,2,*}},
    \mbox{Yuanfeng Ji\textsuperscript{3,7,*}},
    \mbox{Junzhi Ning\textsuperscript{4,*}},
    \mbox{Jiyao Liu\textsuperscript{1,*}},
    \mbox{Yujie Wei\textsuperscript{1}},
    \mbox{Shaohao Rui\textsuperscript{2,6}},
    \mbox{Chaoyang Zhang\textsuperscript{2}},
    \mbox{Wenjie Li\textsuperscript{2,6}},
    \mbox{Wenhao Tang\textsuperscript{4}},
    \mbox{Cheng Tang\textsuperscript{4}},
    \mbox{Wei Li\textsuperscript{4,6}},
    \mbox{Shujian Gao\textsuperscript{1,4}},
    \mbox{Yujie Zhang\textsuperscript{1,2}},
    \mbox{Ying Chen\textsuperscript{4}},
    \mbox{Zilong Li\textsuperscript{1}},
    \mbox{Zongyuan Ge\textsuperscript{5}},
    \mbox{Diping Song\textsuperscript{4}},
    \mbox{Jin Ye\textsuperscript{4,5}},
    \mbox{Ming Hu\textsuperscript{4,5}},
    \mbox{Junjun He\textsuperscript{2,4,7,\textdagger}},
    \mbox{Hongming Shan\textsuperscript{1,\textdagger}}
    \par\vspace{0.6em}
    \normalfont\small
    \mbox{\textsuperscript{1}Fudan University, China}\hspace{1em}%
    \mbox{\textsuperscript{2}Shanghai Innovation Institute, China}\hspace{1em}%
    \mbox{\textsuperscript{3}Stanford University School of Medicine, United States}\hspace{1em}%
    \mbox{\textsuperscript{4}Shanghai Artificial Intelligence Laboratory, China}\hspace{1em}%
    \mbox{\textsuperscript{5}Monash University, Australia}\hspace{1em}%
    \mbox{\textsuperscript{6}Shanghai Jiao Tong University, China}\hspace{1em}%
    \mbox{\textsuperscript{7}Ant Healthcare, Ant Group, China}\par
    \vspace{0.6em}
    \mbox{\texttt{clma24@m.fudan.edu.cn}\hspace{2em}\texttt{yfj@stanford.edu}}\par
  }%
}
\author{\paperauthors}

\begin{document}

\maketitle

\begingroup
\renewcommand{\thefootnote}{\fnsymbol{footnote}}
\footnotetext[1]{These authors contributed equally to this work.}
\footnotetext[2]{Co-corresponding authors. Hongming Shan is affiliated with the Institute of Science and Technology for Brain-inspired Intelligence (ISTBI), the MOE Frontiers Center for Brain Science, and the Key Laboratory of Computational Neuroscience and Brain-Inspired Intelligence (Ministry of Education), Fudan University, China.}
\endgroup

\begin{abstract}
Multimodal medical AI requires a common visual interface for encoding heterogeneous imaging evidence for patient condition modeling. 
However, existing medical visual tokenizers are often tied to a specific dimension and imaging modality, forcing multimodal systems to rely on \emph{fragmented and poorly reusable} representations. 
We introduce \methodname, a general-purpose visual tokenizer that maps 2D images and 3D volumes from diverse imaging modalities into a unified token space. 
Building such a tokenizer poses two key challenges: preserving anatomical structure across dimensions and capturing clinical semantics across modalities. 
To this end, \methodname introduces (\textbf{i}) slice differential regularization, which explicitly models inter-slice anatomical coherence missing in previous per-pixel/per-voxel supervision, and (\textbf{ii}) multi-expert representation alignment, which integrates knowledge from multiple medical expert encoders while retaining modality-specific complementary knowledge. 
Trained on 28M images and 70K volumes across 9 imaging modalities, \methodname supports classification, retrieval, segmentation, synthesis, visual question answering, and medical report generation through a \emph{single} tokenizer. 
Across 24 benchmarks, \methodname achieves state-of-the-art performance, demonstrating a scalable token interface that unifies medical image dimension, modality, and task usage. 
Project page: \url{https://masaaki-75.github.io/medvtok}
\end{abstract}

\begin{figure}[ht]
    \centering
    \includegraphics[width=\linewidth]{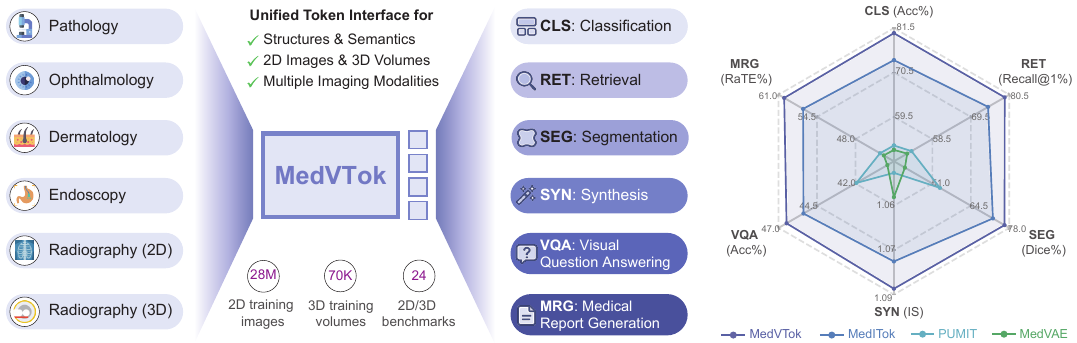}
    \caption{
    \textbf{\methodname provides a unified token interface for multimodal modeling with medical imaging data.} Trained on 28M images and 70K volumes, \methodname achieves state-of-the-art performance across a broad range of downstream tasks covering 24 2D/3D benchmarks.
    }
    \label{fig:teaser}
\end{figure}

\section{Introduction}\label{sec:intro}

Multimodal medical AI aims to integrate heterogeneous patient evidence, including laboratory measurements, longitudinal records, and medical images, to form a holistic understanding of a patient's condition.
As the field shifts from isolated task-specific models toward generalist multimodal systems~\citep{li2026gmai,lin2025healthgpt,ning2026unimedvl}, a central requirement is \emph{a shared interface through which diverse clinical evidence can be encoded, transferred, and reused across tasks}. 
Among these evidence sources, medical images are particularly rich and high-dimensional, capturing anatomical structures, pathological findings, and disease progression. 
Visual tokenization therefore serves as a critical front end that converts raw imaging data into compact visual tokens compatible with multimodal models.

However, medical imaging data are highly heterogeneous across both modality and dimension, including 2D images such as dermoscopic images, color fundus photographs (CFP) and histopathology patches, as well as 3D volumes such as computed tomography (CT) and magnetic resonance imaging (MRI). 
These data differ in acquisition protocol, anatomical coverage, and diagnostic content. 
Existing medical visual tokenizers are often tied to a specific modality and dimension, producing fragmented representations that are difficult to reuse in a multimodal system. 
Consequently, downstream models must absorb much of the burden of modality adaptation and cross-modality representation learning~\citep{qu2024tokenflow}. 
A general-purpose medical visual tokenizer should instead provide a single token interface across modalities while natively supporting both 2D images and 3D volumes.

Building such a tokenizer poses two key challenges. 
\textit{(i) Preserving structural fidelity across dimensions.} Unlike 2D images, 3D scans present volumetric continuity, where organs, vessels, lesions, and tissue boundaries extend across adjacent slices. 
Existing tokenizers often specialize in either 2D or 3D settings~\cite{varma2025medvae,ma2025meditok}, or are optimized for local appearance fidelity or per-pixel/per-voxel reconstruction~\citep{esser2021taming,luo2023pumit}, failing to encode inter-slice structure.
\textit{(ii) Capturing multimodal semantics across modalities.}
Different modalities may describe related clinical concepts while exposing distinct evidence, \eg, histopathology reveals cellular and tissue-level morphology, whereas CT and MRI capture organ-level anatomy.
Existing semantic alignment strategies typically learn representations of different imaging modalities from a single pretrained semantic encoder~\citep{lu2025atoken,ma2025unitok,ma2025meditok,zhang2026both}, which may underrepresent modality-specific semantics outside that encoder's representation space.

In this paper, we introduce \methodname, a general-purpose medical visual tokenizer that maps heterogeneous 2D images and 3D volumes from diverse imaging modalities into a unified token space. 
To preserve structural fidelity across dimensions, \methodname introduces \textit{Slice Differential Regularization} (SliDR) in addition to per-pixel/per-voxel supervision, which explicitly models inter-slice volumetric continuity. 
To capture clinical semantics across modalities, \methodname proposes \textit{Multi-Expert Representation Alignment} (MERA), which builds a shared representation space integrating knowledge from multiple modality-specialized experts instead of a single one, while retaining complementary semantics in each expert's representation space. 
Together, they enable \methodname to produce visual tokens that preserve structural fidelity and encode multimodal clinical semantics.

To train and evaluate \methodname at scale, we curate a large-scale medical imaging corpus spanning 9 modalities, with over 28M 2D images and 70K 3D volumes. 
We assess \methodname across 6 task types: classification, segmentation, retrieval, synthesis, visual question answering, and medical report generation, covering broad utility across 24 benchmarks. 
Our results show that \methodname achieves state-of-the-art performance, demonstrating the value of a scalable token interface that unifies medical image dimension, modality, and downstream task usage.

\noindent\textbf{Contributions.} Our main contributions are as follows. 
\textbf{\textit{(i)}} We introduce \methodname, a general-purpose medical visual tokenizer that provides a shared token interface across diverse imaging modalities and natively supports both 2D images and 3D volumes.
\textbf{\textit{(ii)}} We propose a multi-level supervision strategy that combines slice differential regularization for inter-slice continuity with multi-expert representation alignment for cross-modality clinical semantics.
\textbf{\textit{(iii)}} We curate a large-scale medical imaging corpus and conduct comprehensive evaluations across 24 benchmarks, demonstrating the broad utility of \methodname across classification, segmentation, retrieval, synthesis, visual question answering, and medical report generation.

\section{Related Work}\label{sec:related}

\paragraph{Tokenization across 2D and 3D medical imaging.}
Visual tokenizers convert images into compact tokens while preserving structural fidelity~\citep{van2017vqvae,esser2021taming}.
Recent tokenizers~\citep{wang2024omnitokenizer,lu2025atoken} further explore how a single tokenizer can handle heterogeneous visual formats such as images and videos.
In medical imaging, however, tokenizers are commonly tied to specific input dimensionalities~\citep{varma2025medvae,ma2025meditok}. A few works have explored adaptation to different dimensionalities by pooling 3D weights into 2D~\citep{luo2023pumit} or by duplicating 2D images to fit a 3D model~\citep{wu2025radfm}. While these approaches can handle 2D and 3D inputs simultaneously, they do not consider the respective properties of each dimensionality, \eg, through-plane anatomical continuity in volumes, and thus cannot fully preserve structural fidelity across input forms.
In contrast, \methodname introduces Slice Differential Regularization to explicitly preserve anatomical continuity in volumetric data, and provides a unified 2D--3D token interface via dimension-agnostic patchification.

\paragraph{Semantic representation across medical imaging modalities.}
Medical image representation learning has produced both broad biomedical encoders and modality-specialized expert models.
General medical encoders~\citep{zhang2023biomedclip,codella2024medimageinsight,sellergren2025medgemma,sun2025bmclongclip,khattak2024unimedclip} learn transferable representations from large-scale biomedical data, offering broad coverage across medical domains but potentially missing specialty-specific cues.
Meanwhile, specialized encoders capture finer clinical semantics within individual imaging modalities or specialties~\citep{lu2024conch,yan2025derm1m,du2024retclip,wang2025retizero,yuan2024peskavlp,wu2025radfm,claessens2025spectre,zhao2026hlip}, yet remain isolated in different representation spaces.
Existing tokenizers~\cite{varma2025medvae,ma2025meditok} have only partially captured these different semantics, \eg, MedITok~\citep{ma2025meditok} aligns its representations with those of a single semantic encoder~\citep{zhang2023biomedclip}, thereby inheriting the single-expert bias and limiting the learning of complementary knowledge across modalities.
\methodname instead uses Multi-Expert Representation Alignment to integrate knowledge from multiple medical experts into a single token space, while retaining modality-specific semantics.

\begin{figure}[ht]
    \centering
    \includegraphics[width=\linewidth]{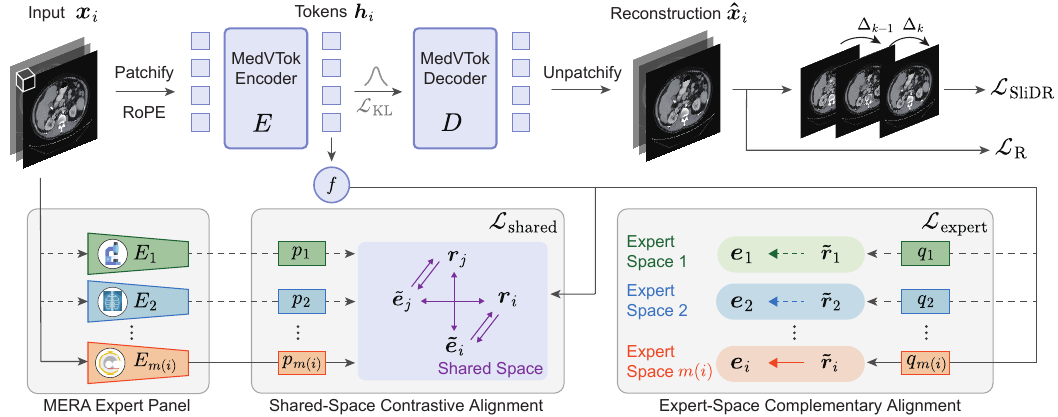}
    \caption{\textbf{Overview of the \methodname training framework.} \methodname learns both (1) structural fidelity across 2D and 3D inputs under the reconstruction loss $\mathcal{L}_\text{R}$ plus the slice differential regularization $\mathcal{L}_\text{SliDR}$, and (2) clinical semantics across modalities driven by $\mathcal{L}_\text{MERA}=\mathcal{L}_\text{shared}{+}\mathcal{L}_\text{expert}$.}
    \label{fig:method}
\end{figure}

\section{Method}\label{sec:method}
\subsection{Overview of \methodname}
\methodname learns a unified latent space that captures both structural fidelity and clinical semantics for 2D and 3D data via two complementary supervision mechanisms: (1) Slice Differential Regularization, which preserves volumetric anatomical continuity beyond what per-voxel reconstruction can capture (\S\ref{sec:stc}), and (2) Multi-Expert Representation Alignment (MERA), which integrates clinical knowledge from multiple modality-specific expert encoders into the shared token space (\S\ref{sec:mera}).

Fig.~\ref{fig:method} illustrates the training framework. Specifically, given an input image or volume $\vct{x} \in \mathbb{R}^{C \times T \times H \times W}$, with $T{=}1$ for images and $T{>}1$ for volumes, \methodname first partitions it into $N$ non-overlapping patches of size $C {\times} P_t {\times} P_h {\times} P_w$, where $N{=} \frac{THW}{P_t P_h P_w}$.
Patches from different samples in a batch are joined into a sequence, where each patch is associated with a coordinate that records its position within the original input. This enables variable-resolution and mixed-dimension batching.

Then, the transformer encoder $E$, with rotary position embeddings (RoPE)~\cite{su2024rope} derived from the coordinates, maps the sequence to per-patch tokens $\vct{h}=E(\vct{x}) \in \mathbb{R}^{N \times d}$.
These tokens are, on the one hand, regularized by a KL divergence term $\mathcal{L}_\text{KL}$ to impose a standard Gaussian constraint after projection and reparameterization; on the other hand, they are processed by the decoder $D$ to reconstruct the original input $\hat{\vct{x}}$.
The reconstruction is supervised by the reconstruction loss $\mathcal{L}_\text{R}$ and the proposed slice differential regularization $\mathcal{L}_\text{SliDR}$.
In parallel, an aggregation head $f$ aggregates the tokens $\vct{h}$ into a holistic representation $\vct{r} = f(\vct{h}) \in \mathbb{R}^{d}$, \footnote{We find that, rather than implementing $f$ as the attention pooling used in prior work~\cite{tschannen2025siglip,lu2025atoken}, adopting simple mean pooling with a linear layer improves dense tasks while retaining performance on other tasks (\S\ref{sec:exp:ablation}).} which is used for semantic alignment via MERA.

\subsection{Slice Differential Regularization}\label{sec:stc}
In tomographic volumes such as CT and MRI scans, adjacent slices share strong anatomical context: organs span continuously across the depth axis, and their boundaries shift gradually from slice to slice.
Per-voxel reconstruction losses like $\ell_1$ can produce outputs that are individually accurate yet exhibit through-plane discontinuities, compromising diagnostic utility.
We address this with the \textit{Slice Differential Regularization}, which operates on inter-slice continuity rather than individual voxels.

For a 3D volume with $T{>}2$ slices along the depth axis, let $\vct{x}_{(k)}$ and $\hat{\vct{x}}_{(k)}$ denote the $k$-th axial slice of the input and the reconstruction, respectively.
We define the first- and second-order slice differentials $\Delta_k(\vct{x}) = \vct{x}_{(k+1)} - \vct{x}_{(k)}$ and $\Delta^2_k(\vct{x}) = \vct{x}_{(k+1)} - 2\vct{x}_{(k)} + \vct{x}_{(k-1)}$, and penalize discrepancies between input and reconstructed transitions:
\begin{equation}\label{eq:stc}
    \mathcal{L}_\text{SliDR} = \frac{1}{T{-}1} \sum_{k=1}^{T-1} \left\| \Delta_k(\hat{\vct{x}}) - \Delta_k(\vct{x}) \right\|_1
    + \frac{1}{T{-}2} \sum_{k=2}^{T-1} \left\| \Delta^2_k(\hat{\vct{x}}) - \Delta^2_k(\vct{x}) \right\|_1,
\end{equation}
where the first term matches inter-slice transitions and the second preserves through-plane curvature.

\subsection{Multi-Expert Representation Alignment}\label{sec:mera}
Apart from structural fidelity, a unified medical visual tokenizer should also encode high-level clinical semantics. 
Achieving this across the full breadth of medical imaging is challenging as different modalities embed distinct knowledge. 
Aligning to a single expert~\citep{zhang2023biomedclip,ma2025meditok} constrains the tokenizer to inherit a single semantic geometry, which may underrepresent modalities outside the expert's training distribution.
We therefore propose \textit{Multi-Expert Representation Alignment} (MERA), which uses a panel of pretrained modality-specific encoders as semantic \textit{experts}. MERA involves two objectives: (i) shared-space contrastive alignment, which jointly aligns representations from multiple modality experts into a shared space, and (ii) expert-space complementary alignment, which preserves complementary information in each modality expert's representation space.

\paragraph{Shared-space contrastive alignment.} 
An input $\vct{x}_i$ of imaging modality $m(i)$ is routed to the corresponding expert $E_{m(i)}$ to produce a feature $\vct{e}_i = E_{m(i)}(\vct{x}_i)$, which will be linearly projected to a shared space, giving $\tilde{\vct{e}}_{i} = p_{m(i)}(\vct{e}_i)\in\mathbb{R}^{d}$.
We then apply a contrastive loss $\mathcal{L}_\text{shared}$ between \methodname's representations $\vct{r}_i$ and expert features $\tilde{\vct{e}}_{i}$ within a batch with a learnable temperature $\tau$:
\begin{equation}\label{eq:mera-sec}
    \mathcal{L}_\text{shared} = -\frac{1}{2B} \sum_{i=1}^{B} \left[
    \log \frac{\exp(\vct{r}_i \cdot \tilde{\vct{e}}_{i}/\tau)}{\sum_{j=1}^{B} \exp(\vct{r}_i \cdot \tilde{\vct{e}}_{j}/\tau)}
    +\log \frac{\exp\bigl(\vct{r}_i \cdot \tilde{\vct{e}}_{i} / \tau \bigr)}{\sum_{j=1}^{B} \exp\bigl(\vct{r}_{j} \cdot \tilde{\vct{e}}_i / \tau\bigr)}
    \right].
\end{equation}

\paragraph{Expert-space complementary alignment.} Note that $\mathcal{L}_\text{shared}$ projects heterogeneous expert features into a single space, which may discard modality-specific nuances present in each expert's native space. To address this issue, we introduce expert-space complementary learning. 
For each modality $m(i)$, a backprojector $q_{m(i)}$ maps $\vct{r}_i$ back to the expert's space, yielding $\tilde{\vct{r}}_i=q_{m(i)}(\vct{r}_i)\in\mathbb{R}^{d_{m(i)}}$. We then enforce cosine similarity between backprojected representations and original expert features:
\begin{equation}\label{eq:mera-nec}
    \mathcal{L}_\text{expert} = \frac{1}{B} \sum_{i=1}^{B} \left(1 - \frac{\tilde{\vct{r}}_i \cdot \vct{e}_i}{\|\tilde{\vct{r}}_i\| \cdot \|\vct{e}_i\|}\right).
\end{equation}
Together, $\mathcal{L}_\text{shared}$ and $\mathcal{L}_\text{expert}$ provide bidirectional representation alignment: the former anchors \methodname within a single representation space, while the latter preserves modality-specific expert knowledge that might otherwise be compressed away. The total loss for MERA is given by:
\begin{equation}\label{eq:mera}
    \mathcal{L}_\text{MERA} = \mathcal{L}_\text{shared} + \mathcal{L}_\text{expert}.
\end{equation}

\subsection{Training and Inference}\label{sec:training}
\methodname is trained with the combined objective:
\begin{equation}\label{eq:total_loss}
    \mathcal{L} = \mathcal{L}_\text{R}
    + \lambda_\text{KL}\mathcal{L}_\text{KL}
    + \lambda_\text{SliDR} \mathcal{L}_\text{SliDR}
    + \lambda_\text{MERA} \mathcal{L}_\text{MERA},
\end{equation}
where the standard~\cite{esser2021taming,rombach2022high,ma2025unitok} reconstruction loss $\mathcal{L}_\text{R}$ consists of pixel-level, LPIPS~\cite{zhang2018unreasonable}, and adversarial losses.
Training proceeds in two stages: Stage~1 uses only 2D images to build foundational visual structure and semantic capabilities, with $\lambda_\text{SliDR}=0$; Stage~2 continues on both 2D images and 3D volumes with $\lambda_\text{SliDR}=0.1$.
At inference, \methodname serves as a general-purpose tokenizer: given an image or volume, it encodes tokens that are directly consumed by downstream models and transferable across various tasks without further fine-tuning of the tokenizer.

\section{Experiments}\label{sec:exp}

\subsection{Pretraining}\label{sec:exp:pretrain}
\paragraph{Training data.} We prepare the training data by aggregating existing large-scale publicly available training-ready datasets to ensure image quality and diversity. In total, we obtain 28,769,502 images and 70,144 volumes for training \methodname. Please refer to \S~\ref{sec:app:training_datasets} for data statistics.

\paragraph{Implementation details.}
We use the AdamW~\cite{loshchilov2018adamw} optimizer with $(\beta_1, \beta_2) = (0.9, 0.95)$ and a cosine annealing schedule for 100K steps per stage. The learning rate decays from $10^{-4}$ to $10^{-5}$ in Stage~1 and from $2{\times}10^{-5}$ to $2{\times}10^{-6}$ in Stage~2, with $\lambda_\text{KL}=10^{-6}$, $\lambda_\text{MERA}=1$, and $\lambda_\text{SliDR}=0.1$. During Stage~2 training, 2D and 3D data are fed into \methodname in alternating iterations, with an effective batch size of 600 for 2D images and 48 for 3D volumes. Please see \S~\ref{sec:app:training_setup} for details.

\subsection{Evaluation on Downstream Tasks}\label{sec:exp:downstream}

\paragraph{Evaluation protocol.} We compare \methodname with both medical-domain tokenizers~\cite{ma2025meditok,varma2025medvae,luo2023pumit} and general-domain ones~\cite{rombach2022high,qu2024tokenflow,huang2025mingunivision,ma2025unitok,lu2025atoken}. We first evaluate token fidelity of different tokenizers via autoencoding reconstruction, and then evaluate token representation quality on 6 downstream tasks: image classification, image-to-image retrieval, image segmentation, image synthesis, visual question answering (VQA), and medical report generation. \emph{All evaluations use a frozen tokenizer and focus on the quality of the encoded tokens rather than building task-specific models} (\S~\ref{sec:app:eval_setup}).

\paragraph{Evaluation data.} The token fidelity, classification, and retrieval evaluations share 9 benchmarks: JSIEC (fundus, retinal conditions)~\cite{cen2021automatic}, ChestCTScan (CT, chest cancers)~\cite{syed2024medicalimagining}, HAM10K (dermoscopy, skin diseases)~\cite{tschandl2018ham10000}, BRISC (MRI, brain tumor)~\cite{fateh2025brisc}, PolypDiag (endoscopy, polyp detection)~\cite{tian2022contrastive}, OCTDL (OCT, retinal diseases)~\cite{kulyabin2024octdl}, LungHist700 (pathology, lung cancers)~\cite{diosdado2024lunghist700}, BUSBRA (ultrasound, tumor staging)~\cite{gomez2024bus}, and Kermany (X-ray, pneumonia)~\cite{kermany2018identifying}.
Segmentation datasets include MaplesDR (fundus, optic disc/cup)~\cite{lepetit2024maples}, MosMedDataPlus (CT, COVID-19 lesion)~\cite{morozov2020mosmeddata}, HAM10K (dermoscopy, lesion)~\cite{tschandl2018ham10000}, PolypDB (endoscopy, polyp)~\cite{jha2024polypdb}, BRISC (MRI, brain tumor)~\cite{fateh2025brisc}, PanNuke (pathology, different cells)~\cite{gamper2019pannuke}, BUSBRA (ultrasound, tumor)~\cite{gomez2024bus}, and VinDr-RibCXR (X-ray, rib)~\cite{nguyen2021vindr}.
Image synthesis uses images from the GMAI-VL-5.5M~\cite{li2026gmai} dataset. VQA benchmarks include GMAI-MMBench~\cite{chen2024gmai}, Fundus-MMBench~\cite{liu2025constructing}, PathVQA~\cite{he2020pathvqa}, MM-Skin~\cite{zeng2025mm}, Kvasir-VQA-x1~\cite{gautam2025kvasir}, and SLAKE~\cite{liu2021slake}. For 3D tasks, we use AMOS-MM~\cite{ji2022amos} and MRISegmenter~\cite{zhuang2025mrisegmenter} to evaluate 3D token fidelity. AMOS-MM CT--report pairs are used for report generation evaluation (\S~\ref{sec:app_eval_data}).

\begin{table*}[h]
\centering
\caption{\textbf{Token fidelity evaluation via autoencoding reconstruction} [PSNR (dB)$\uparrow$, SSIM (\%)$\uparrow$]. 3D columns are excluded from the 2D average. MedVAE uses separate models for 2D and 3D inputs. The best result is marked in \textbf{bold} and the second-best is \underline{underlined}.}
\label{tab:rec}
\resizebox{\textwidth}{!}{
\begin{tabular}{llcccccccccc|cc}
\toprule
Metrics & Method & CFP & CT & Derm. & Endo. & MRI & OCT & Path. & US & X-ray & Avg. & 3D-MRI & 3D-CT \\
\midrule
 \multirow{9}{*}{PSNR ($\uparrow$)} & SDVAE~\citep{rombach2022high} & 37.62 & 36.38 & 34.68 & 33.57 & 34.31 & 29.17 & 24.43 & 32.13 & 35.29 & 33.06 & -- & -- \\
  & TokenFlow~\citep{qu2024tokenflow} & 30.57 & 25.45 & 29.54 & 25.75 & 24.51 & 24.79 & 20.17 & 22.82 & 26.48 & 25.56 & -- & -- \\
  & MingTok~\citep{huang2025mingunivision} & 35.58 & 31.23 & 32.43 & 29.85 & 29.37 & 26.92 & 21.87 & 26.42 & 31.25 & 29.43 & -- & -- \\
  & UniTok~\citep{ma2025unitok} & 34.02 & 31.32 & 31.89 & 30.34 & 30.82 & 28.00 & 24.40 & 29.48 & 31.57 & 30.21 & -- & -- \\
  & AToken~\citep{lu2025atoken} & \underline{41.08} & \underline{41.60} & \underline{39.00} & \underline{39.69} & \underline{41.64} & \underline{32.94} & \underline{28.57} & \underline{39.41} & \underline{40.02} & \underline{38.22} & \underline{31.46} & \underline{35.68} \\
  \cmidrule(lr){2-14}
  & PUMIT~\citep{luo2023pumit} & 32.03 & 32.86 & 27.05 & 28.73 & 32.88 & 27.06 & 19.43 & 23.26 & 26.69 & 27.78 & 29.96 & 31.87 \\
  & MedVAE~\citep{varma2025medvae} & 17.11 & 37.97 & 22.42 & 16.83 & 35.72 & 30.90 & 20.13 & 33.47 & 37.04 & 27.95 & 25.89 & 31.30 \\
  & MedITok~\citep{ma2025meditok} & 35.92 & 35.61 & 33.81 & 32.42 & 34.82 & 30.26 & 26.03 & 32.62 & 34.70 & 32.91 & -- & -- \\
  & \cellcolor{gray!20}\methodname{} & \cellcolor{gray!20}\textbf{42.22} & \cellcolor{gray!20}\textbf{44.75} & \cellcolor{gray!20}\textbf{40.39} & \cellcolor{gray!20}\textbf{41.42} & \cellcolor{gray!20}\textbf{43.75} & \cellcolor{gray!20}\textbf{35.04} & \cellcolor{gray!20}\textbf{30.16} & \cellcolor{gray!20}\textbf{42.08} & \cellcolor{gray!20}\textbf{41.83} & \cellcolor{gray!20}\textbf{40.18} & \cellcolor{gray!20}\textbf{31.77} & \cellcolor{gray!20}\textbf{36.50} \\
\midrule
 \multirow{9}{*}{SSIM ($\uparrow$)} & SDVAE~\citep{rombach2022high} & 89.99 & 90.47 & 85.73 & 90.87 & 91.53 & 64.05 & 53.85 & 86.91 & 90.24 & 82.63 & -- & -- \\
  & TokenFlow~\citep{qu2024tokenflow} & 83.75 & 73.40 & 78.03 & 77.23 & 74.13 & 59.89 & 31.75 & 55.34 & 76.99 & 67.84 & -- & -- \\
  & MingTok~\citep{huang2025mingunivision} & 88.63 & 83.66 & 79.14 & 83.83 & 83.49 & 55.11 & 38.00 & 66.76 & 83.88 & 73.61 & -- & -- \\
  & UniTok~\citep{ma2025unitok} & 88.54 & 88.70 & 82.48 & 89.62 & 90.12 & 65.24 & 55.36 & 87.21 & 88.20 & 81.72 & -- & -- \\
  & AToken~\citep{lu2025atoken} & \underline{94.74} & \underline{96.24} & \underline{92.51} & \underline{96.98} & \underline{97.61} & \underline{80.60} & \underline{74.24} & \underline{97.13} & \underline{95.42} & \underline{91.72} & 85.19 & \underline{95.50} \\
  \cmidrule(lr){2-14}
  & PUMIT~\citep{luo2023pumit} & 88.09 & 90.82 & 84.84 & 90.17 & 90.92 & 65.34 & 57.92 & 81.99 & 89.30 & 82.16 & \underline{85.20} & 90.55 \\
  & MedVAE~\citep{varma2025medvae} & 77.26 & 92.67 & 86.63 & 83.86 & 93.87 & 72.44 & 54.18 & 89.97 & 92.90 & 82.64 & 75.14 & 90.44 \\
  & MedITok~\citep{ma2025meditok} & 91.50 & 92.74 & 88.22 & 93.47 & 95.13 & 73.81 & 64.52 & 92.95 & 92.46 & 87.20 & -- & -- \\
  & \cellcolor{gray!20}\methodname{} & \cellcolor{gray!20}\textbf{95.95} & \cellcolor{gray!20}\textbf{97.86} & \cellcolor{gray!20}\textbf{94.51} & \cellcolor{gray!20}\textbf{97.90} & \cellcolor{gray!20}\textbf{98.66} & \cellcolor{gray!20}\textbf{86.54} & \cellcolor{gray!20}\textbf{80.52} & \cellcolor{gray!20}\textbf{98.28} & \cellcolor{gray!20}\textbf{97.21} & \cellcolor{gray!20}\textbf{94.16} & \cellcolor{gray!20}\textbf{87.79} & \cellcolor{gray!20}\textbf{95.65} \\
\bottomrule
\end{tabular}
}
\end{table*}

\begin{figure}[b]
    \centering
    \includegraphics[width=\linewidth]{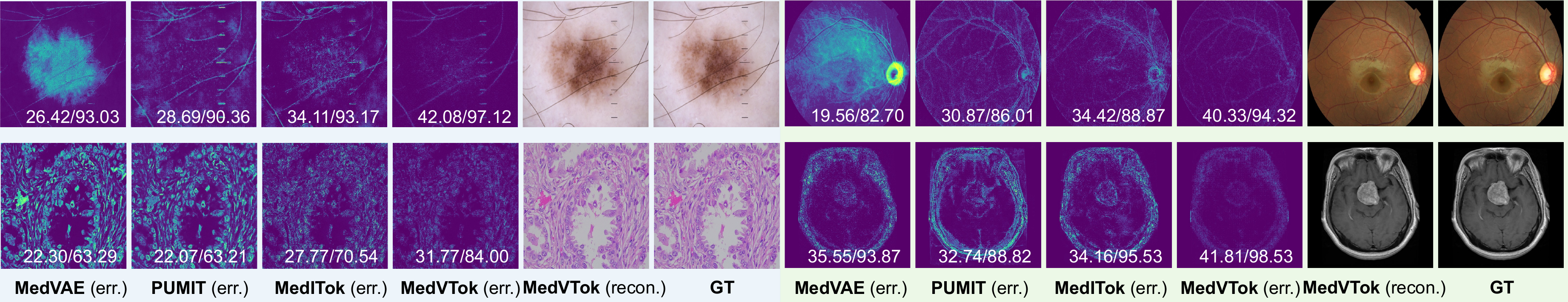}
    \caption{\textbf{Qualitative reconstruction for token fidelity evaluation.} For each example, we show per-tokenizer absolute error maps with PSNR/SSIM values reported per panel.}
    \label{fig:result_recon}
\end{figure}

\paragraph{Token fidelity evaluation.}\label{sec:exp:recon}
We assess token fidelity by how well the tokenizer reconstructs input data from its encoded tokens, reporting PSNR and SSIM between the input and the reconstructed data.
Table~\ref{tab:rec} reports 2D reconstruction quality. \methodname ranks first on all nine modalities in both PSNR and SSIM. The improvement is most notable in challenging domains such as pathology, where other tokenizers suffer from limited training diversity. Medical-domain tokenizers (MedITok, MedVAE) and general-domain ones (TokenFlow, UniTok, MingTok) lag significantly behind, confirming that medical images benefit from specialized architectures and large-scale multimodal training. Fig.~\ref{fig:result_recon} visualizes the reconstruction error maps of the competing methods and the reconstructed images produced by \methodname.
The last two columns of Table~\ref{tab:rec} report native 3D reconstruction. \methodname achieves the best PSNR and SSIM on both MRI and CT volumes by a large margin over PUMIT and MedVAE. The strong 3D performance suggests that our unified 2D--3D training paradigm effectively transfers learned representations to volumetric data. Please see \S~\ref{sec:app:eval_setup} for qualitative comparisons.

\begin{table*}[t]
\centering
\caption{\textbf{Image classification and retrieval evaluation} [Accuracy (\%)$\uparrow$, Recall@1 (\%)$\uparrow$]. *: semantic encoder.}
\label{tab:cls_ret}
\resizebox{\textwidth}{!}{
\begin{tabular}{llcccccccccc}
\toprule
Metrics & Method & CFP & CT & Derm. & Endo. & MRI & OCT & Path. & US & X-ray & Avg. \\
\midrule
 \multirow{13}{*}{\shortstack{Acc. ($\uparrow$)}} 
  & SDVAE~\citep{rombach2022high} & 18.79 & 45.68 & 60.23 & \underline{47.59} & 77.47 & 62.94 & 47.56 & 43.97 & 78.69 & 53.66 \\
  & TokenFlow~\citep{qu2024tokenflow} & 35.73 & 46.10 & 61.95 & 42.83 & 79.73 & 68.53 & 45.38 & 46.10 & 86.91 & 57.03 \\
  & MingTok~\citep{huang2025mingunivision} & 50.71 & 48.10 & 67.86 & 44.93 & 77.70 & 64.39 & 64.87 & 45.04 & 83.65 & 60.81 \\
  & UniTok~\citep{ma2025unitok} & 66.95 & 74.79 & 70.63 & 44.82 & 80.53 & 72.88 & 69.23 & 46.28 & 91.03 & 68.57 \\
  & AToken~\citep{lu2025atoken} & 64.12 & 65.82 & 72.31 & 46.63 & 88.13 & 66.46 & 66.54 & 46.19 & 91.29 & 67.50 \\
  \cmidrule{2-12}
  & BiomedCLIP$^{*}$~\citep{zhang2023biomedclip} & 60.59 & 66.03 & 68.78 & 46.03 & 88.43 & 73.91 & 79.10 & 48.32 & 93.80 & 69.45 \\
  & MedSigLIP$^{*}$~\citep{sellergren2025medgemma} & \underline{81.07} & 73.73 & 72.13 & 46.11 & 91.73 & 75.88 & \textbf{84.10} & 46.99 & 95.30 & 74.12 \\
  & MedImageInsight$^{*}$~\citep{codella2024medimageinsight} & 75.64 & 75.32 & \textbf{85.87} & 44.65 & \textbf{95.47} & \underline{90.79} & 71.79 & \underline{50.18} & \underline{96.21} & \underline{76.21} \\
  & PUMIT~\citep{luo2023pumit} & 13.84 & 43.78 & 60.34 & 45.85 & 78.00 & 59.52 & 47.44 & 43.44 & 79.49 & 52.41 \\
  & MedVAE~\citep{varma2025medvae} & 13.28 & 45.25 & 60.12 & 45.34 & 75.57 & 57.56 & 49.62 & 41.76 & 73.40 & 51.32 \\
  & MedITok~\citep{ma2025meditok} & 74.29 & \underline{78.80} & 72.82 & 46.80 & 91.23 & 78.67 & 74.23 & 50.09 & 95.03 & 73.55 \\
  & \cellcolor{gray!20}\methodname{} & \cellcolor{gray!20}\textbf{90.25} & \cellcolor{gray!20}\textbf{89.77} & \cellcolor{gray!20}\underline{74.16} & \cellcolor{gray!20}\textbf{48.71} & \cellcolor{gray!20}\underline{94.23} & \cellcolor{gray!20}\textbf{94.00} & \cellcolor{gray!20}\underline{82.69} & \cellcolor{gray!20}\textbf{51.33} & \cellcolor{gray!20}\textbf{96.90} & \cellcolor{gray!20}\textbf{80.23} \\
\midrule
 \multirow{13}{*}{\shortstack{Recall@1 ($\uparrow$)}} 
  & SDVAE~\citep{rombach2022high} & 22.75 & 43.81 & 57.41 & \textbf{56.14} & 90.10 & 57.14 & 32.30 & 39.10 & 82.53 & 53.47 \\
  & TokenFlow~\citep{qu2024tokenflow} & 45.06 & 50.79 & 61.38 & 45.08 & 91.00 & 68.63 & 35.02 & 44.41 & 87.50 & 58.76 \\
  & MingTok~\citep{huang2025mingunivision} & 54.94 & 52.38 & 63.16 & 44.82 & 89.50 & 68.01 & 53.70 & 42.29 & 83.01 & 61.31 \\
  & UniTok~\citep{ma2025unitok} & 69.10 & 88.25 & 64.22 & 50.48 & 89.80 & 73.60 & 67.70 & 44.87 & 87.98 & 70.67 \\
  & AToken~\citep{lu2025atoken} & 66.09 & 78.10 & 68.72 & 44.04 & 95.10 & 74.53 & 63.42 & 43.62 & 88.46 & 69.12 \\
  \cmidrule{2-12}
  & BiomedCLIP$^{*}$~\citep{zhang2023biomedclip} & 62.66 & 83.81 & 66.27 & 44.56 & 96.00 & 76.40 & 70.43 & 45.48 & 89.10 & 70.52 \\
  & MedSigLIP$^{*}$~\citep{sellergren2025medgemma} & \underline{82.40} & \textbf{92.06} & 67.79 & 44.82 & \underline{96.90} & 77.02 & \textbf{83.27} & 39.63 & \underline{94.55} & 75.38 \\
  & MedImageInsight$^{*}$~\citep{codella2024medimageinsight} & 78.54 & 86.67 & \textbf{84.13} & 44.30 & \textbf{97.80} & \underline{92.86} & 70.04 & \underline{46.54} & \textbf{95.19} & \underline{77.34} \\
  & PUMIT~\citep{luo2023pumit} & 15.88 & 49.52 & 58.66 & 48.96 & 89.80 & 51.24 & 38.52 & 36.70 & 83.49 & 52.53 \\
  & MedVAE~\citep{varma2025medvae} & 12.45 & 47.30 & 56.15 & \underline{53.69} & 88.90 & 52.48 & 38.52 & 35.90 & 76.28 & 51.30 \\
  & MedITok~\citep{ma2025meditok} & 73.82 & 87.62 & 68.65 & 52.52 & 95.60 & 80.75 & 72.76 & 46.01 & 92.15 & 74.43 \\
  & \cellcolor{gray!20}\methodname{} & \cellcolor{gray!20}\textbf{90.99} & \cellcolor{gray!20}\underline{90.16} & \cellcolor{gray!20}\underline{70.70} & \cellcolor{gray!20}52.59 & \cellcolor{gray!20}96.10 & \cellcolor{gray!20}\textbf{93.79} & \cellcolor{gray!20}\underline{75.49} & \cellcolor{gray!20}\textbf{47.61} & \cellcolor{gray!20}\textbf{95.19} & \cellcolor{gray!20}\textbf{79.18} \\
\bottomrule
\end{tabular}
}
\end{table*}

\paragraph{Image classification and retrieval evaluation.}\label{sec:exp:clsret}
To directly measure the representation quality for image classification, we adopt a $k$-nearest neighbor protocol. Each test sample is classified based on its $k$ nearest training neighbors with $k \in \{10, 20, 100, 200\}$, and we report the highest accuracy among $k$ values for each tokenizer.
For retrieval, the training split serves as the gallery, and the test split provides the queries. Gallery images are ranked by cosine similarity, and we report Recall@1 performance.

Results are presented in Table~\ref{tab:cls_ret}. \methodname achieves the highest average on both tasks, surpassing not only other tokenizers but also dedicated semantic encoders (BiomedCLIP, MedImageInsight, MedSigLIP) that are explicitly trained for representation learning. This indicates that the features produced by \methodname are highly discriminative. The consistent ranking further confirms that the learned representations generalize across different downstream protocols.

\begin{table*}[t]
\centering
\caption{\textbf{Image segmentation evaluation} [Dice (\%)$\uparrow$]. *: semantic encoder.}
\label{tab:seg}
\resizebox{\textwidth}{!}{
\begin{tabular}{llccccccccc}
\toprule
Metrics & Method & CFP & CT & Derm. & Endo. & MRI & Path. & US & X-ray & Avg. \\
\midrule
 \multirow{13}{*}{\shortstack{Dice ($\uparrow$)}}
  & SDVAE~\citep{rombach2022high} & 73.03 & 14.88 & 62.37 & 24.67 & 12.69 & 49.11 & 84.19 & 35.56 & 44.56 \\
  & TokenFlow~\citep{qu2024tokenflow} & 70.39 & 19.05 & 72.62 & 40.53 & 30.49 & 52.34 & 91.33 & 49.52 & 53.28 \\
  & MingTok~\citep{huang2025mingunivision} & 85.36 & 27.03 & 83.91 & 57.32 & 52.54 & 67.68 & 92.91 & 54.71 & 65.18 \\
  & UniTok~\citep{ma2025unitok} & 60.93 & 33.13 & 86.25 & 67.25 & 54.90 & 72.51 & 93.35 & 57.60 & 65.74 \\
  & AToken~\citep{lu2025atoken} & 97.09 & 37.33 & 87.45 & \underline{69.88} & 63.57 & 74.30 & 94.90 & 59.69 & 73.03 \\
  \cmidrule{2-11}
  & BiomedCLIP$^{*}$~\citep{zhang2023biomedclip} & 50.22 & 31.83 & 87.48 & 60.64 & 57.95 & 73.65 & 95.41 & 59.83 & 64.63 \\
  & MedSigLIP$^{*}$~\citep{sellergren2025medgemma} & \underline{97.28} & \underline{40.79} & \underline{87.76} & 69.81 & \underline{67.32} & \underline{76.20} & 95.33 & \textbf{62.48} & \underline{74.62} \\
  & MedImageInsight$^{*}$~\citep{codella2024medimageinsight} & 92.34 & 25.60 & 85.45 & 66.96 & 55.81 & 42.93 & \underline{95.80} & 50.16 & 64.38 \\
  & PUMIT~\citep{luo2023pumit} & 40.05 & 25.53 & 77.94 & 37.32 & 36.69 & 66.00 & 90.73 & 54.82 & 53.63 \\
  & MedVAE~\citep{varma2025medvae} & 61.74 & 12.57 & 57.59 & 19.73 & 10.13 & 36.40 & 81.90 & 30.81 & 38.86 \\
  & MedITok~\citep{ma2025meditok} & 96.70 & 40.52 & 80.28 & 66.07 & 62.42 & 75.80 & 94.65 & 59.47 & 71.99 \\
  & \cellcolor{gray!20}\methodname{} & \cellcolor{gray!20}\textbf{97.44} & \cellcolor{gray!20}\textbf{42.91} & \cellcolor{gray!20}\textbf{90.25} & \cellcolor{gray!20}\textbf{72.52} & \cellcolor{gray!20}\textbf{69.62} & \cellcolor{gray!20}\textbf{77.74} & \cellcolor{gray!20}\textbf{96.91} & \cellcolor{gray!20}\underline{61.13} & \cellcolor{gray!20}\textbf{76.07} \\
\bottomrule
\end{tabular}
}
\end{table*}

\paragraph{Image segmentation evaluation.}\label{sec:exp:seg}
We follow Sim{\'e}oni \etal~\cite{simeoni2025dinov3} and adopt a linear segmentation protocol to evaluate the representation quality for image segmentation. Specifically, a single linear segmentation head is trained with a cross-entropy loss on top of the frozen patch tokens produced by each tokenizer, with learning rates searched over $\{10^{-2}, 10^{-3}, 10^{-4}\}$. For each tokenizer, the best-performing head on the validation set is selected for evaluation.
Results are presented in Table~\ref{tab:seg}. \methodname achieves the best average Dice score, outperforming all baselines including semantic encoders. This task reveals a clear hierarchy: although general autoencoding tokenizers (\eg, SDVAE, PUMIT) can embed rich low-level details into the latent space, their high-level semantics are not activated, resulting in low segmentation performance. Semantic encoders provide explicit visual semantics and significantly outperform the autoencoding tokenizers. Unified tokenizers (MedITok, \methodname) encode both fine-grained spatial information and high-level semantics, enjoying the benefits of both and improving dense prediction. Please see \S~\ref{sec:app:eval_setup} for qualitative comparisons.

\begin{figure}[t]
    \centering
    \includegraphics[width=\linewidth]{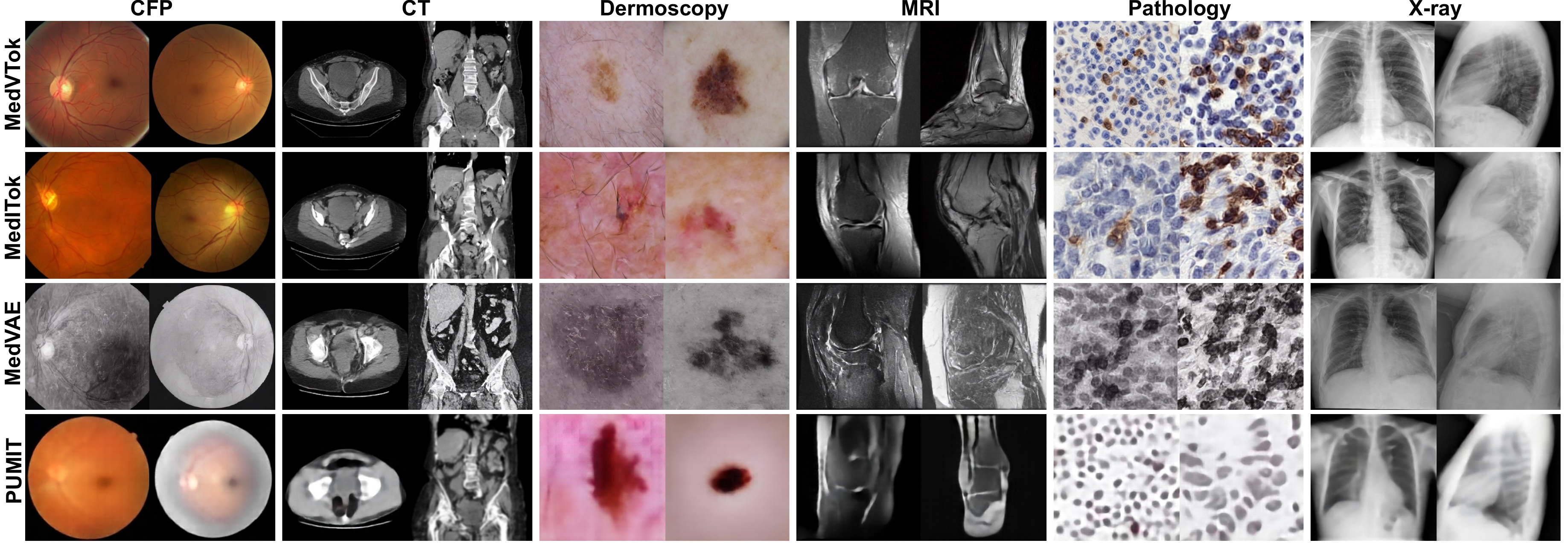}
    \caption{\textbf{Image synthesis examples.} A latent diffusion model trained from scratch on \methodname{}'s tokens produces samples with sharper anatomical structure and fewer artifacts than competing tokenizers across modalities, including fundus, X-ray, and dermoscopy.}
    \label{fig:result_syn}
\end{figure}

\paragraph{Image synthesis evaluation.}\label{sec:exp:gen}
We evaluate how the tokenizer influences text-conditioned medical image synthesis within a latent diffusion~\cite{rombach2022high} framework. A UNet denoising backbone is trained from scratch in each tokenizer's latent space while keeping both the tokenizer and the text encoder frozen. This design isolates the effect of the tokenizer: the architecture, noise scheduler, optimizer, and training data are held constant across experiments.
Training uses 70K image--text pairs spanning seven imaging modalities. 
We report the Fréchet Inception Distance (FID)~\cite{heusel2017fid,woodland2024fidmed} in Table~\ref{tab:syn}; lower FID indicates higher fidelity.
\methodname achieves the highest image synthesis quality with the best average FID, outperforming both specialized tokenizers (PUMIT, MedVAE) and the previous unified tokenizer (MedITok), as shown in Fig.~\ref{fig:result_syn}.
PUMIT and MedVAE consistently underperform, confirming that synthesis quality depends not only on reconstruction fidelity but also on a well-structured, semantically meaningful latent space that facilitates the diffusion model's learning.

\begin{table*}[t]
\centering
\caption{\textbf{Image synthesis evaluation} (FID $\downarrow$). A latent diffusion model is trained from scratch in each frozen tokenizer's latent space with identical architecture and data.}
\label{tab:syn}
\resizebox{\textwidth}{!}{
\setlength{\tabcolsep}{10pt}
\begin{tabular}{lcccccccc}
\toprule
Method & CFP & CT & Derm. & Endo. & MRI & Path. & X-ray & Avg. \\
\midrule
PUMIT~\citep{luo2023pumit} & 140.40 & 183.49 & 272.26 & 191.49 & 227.64 & 377.12 & 200.14 & 227.51 \\
MedVAE~\citep{varma2025medvae} & 118.13 & 154.42 & 202.09 & 170.29 & 145.25 & 325.87 & 88.55 & 172.08 \\
MedITok~\citep{ma2025meditok} & \underline{90.65} & \textbf{142.46} & \underline{176.94} & \textbf{58.38} & \underline{122.96} & \underline{99.66} & \underline{77.24} & \underline{109.76} \\
\cellcolor{gray!20}\methodname{} & \cellcolor{gray!20}\textbf{59.85} & \cellcolor{gray!20}\underline{146.39} & \cellcolor{gray!20}\textbf{142.87} & \cellcolor{gray!20}\underline{82.39} & \cellcolor{gray!20}\textbf{116.78} & \cellcolor{gray!20}\textbf{94.65} & \cellcolor{gray!20}\textbf{69.17} & \cellcolor{gray!20}\textbf{101.73} \\
\bottomrule
\end{tabular}
}
\end{table*}

\begin{table*}[ht]
\centering
\caption{\textbf{Visual question answering evaluation} [Accuracy (\%)$\uparrow$]. Each frozen tokenizer is plugged into the LLaVA-Med~\cite{li2023llavamed} framework for evaluation. %
}
\label{tab:vqa}
\resizebox{\textwidth}{!}{
\begin{tabular}{lccccccc}
\toprule
Method & FundusMM & MM-Skin & Kvasir & PathVQA & SLAKE & GMAI-MM & Avg. \\
\midrule
 PUMIT~\citep{luo2023pumit} & 29.84{\scriptsize$\pm$0.42} & 34.74{\scriptsize$\pm$0.38} & 69.50{\scriptsize$\pm$0.51} & 52.97{\scriptsize$\pm$0.63} & 36.44{\scriptsize$\pm$0.55} & 28.16{\scriptsize$\pm$0.47} & 41.94{\scriptsize$\pm$0.33} \\
 MedVAE~\citep{varma2025medvae} & 28.39{\scriptsize$\pm$0.56} & 34.21{\scriptsize$\pm$0.44} & 66.64{\scriptsize$\pm$0.72} & 51.15{\scriptsize$\pm$0.58} & 31.57{\scriptsize$\pm$0.67} & 27.61{\scriptsize$\pm$0.53} & 39.93{\scriptsize$\pm$0.41} \\
 MedITok~\citep{ma2025meditok} & \underline{31.29}{\scriptsize$\pm$0.35} & \underline{36.18}{\scriptsize$\pm$0.41} & \underline{70.63}{\scriptsize$\pm$0.29} & \textbf{58.46}{\scriptsize$\pm$0.52} & \underline{45.61}{\scriptsize$\pm$0.38} & \underline{30.15}{\scriptsize$\pm$0.26} & \underline{45.39}{\scriptsize$\pm$0.28} \\
 \cellcolor{gray!20}\methodname{} & \cellcolor{gray!20}\textbf{33.39}{\scriptsize$\pm$0.31} & \cellcolor{gray!20}\textbf{37.49}{\scriptsize$\pm$0.28} & \cellcolor{gray!20}\textbf{72.98}{\scriptsize$\pm$0.34} & \cellcolor{gray!20}\underline{57.24}{\scriptsize$\pm$0.45} & \cellcolor{gray!20}\textbf{45.94}{\scriptsize$\pm$0.37} & \cellcolor{gray!20}\textbf{31.81}{\scriptsize$\pm$0.29} & \cellcolor{gray!20}\textbf{46.48}{\scriptsize$\pm$0.22} \\
\bottomrule
\end{tabular}
}
\end{table*}

\begin{table*}[ht]
\centering
\caption{\textbf{Medical report generation evaluation.} Each frozen tokenizer is integrated into the LaMed~\cite{bai2024m3d} framework for evaluation, generating findings in abdomen, chest, and pelvis regions.}
\label{tab:mrg}
\footnotesize
\begin{tabularx}{\textwidth}{ll*{4}{>{\centering\arraybackslash}X}}
\toprule
Metrics & Method & Abdomen & Chest & Pelvis & Avg. \\
\midrule
\multirow{4}{*}{BLEU-2 ($\uparrow$)}
 & PUMIT~\citep{luo2023pumit} & \textbf{0.289}{\scriptsize$\pm$0.008} & 0.193{\scriptsize$\pm$0.003} & 0.164{\scriptsize$\pm$0.007} & 0.215{\scriptsize$\pm$0.005} \\
 & MedVAE-3D~\citep{varma2025medvae} & 0.249{\scriptsize$\pm$0.038} & 0.185{\scriptsize$\pm$0.020} & 0.152{\scriptsize$\pm$0.010} & 0.195{\scriptsize$\pm$0.013} \\
 & MedITok~\citep{ma2025meditok} & 0.256{\scriptsize$\pm$0.013} & \underline{0.219}{\scriptsize$\pm$0.020} & \textbf{0.294}{\scriptsize$\pm$0.017} & \underline{0.257}{\scriptsize$\pm$0.012} \\
 & \cellcolor{gray!20}\methodname{} & \cellcolor{gray!20}\underline{0.272}{\scriptsize$\pm$0.005} & \cellcolor{gray!20}\textbf{0.263}{\scriptsize$\pm$0.045} & \cellcolor{gray!20}\underline{0.288}{\scriptsize$\pm$0.017} & \cellcolor{gray!20}\textbf{0.274}{\scriptsize$\pm$0.009} \\
\midrule
\multirow{4}{*}{RaTEScore ($\uparrow$)}
& PUMIT~\citep{luo2023pumit} & 0.604{\scriptsize$\pm$0.005} & \underline{0.511}{\scriptsize$\pm$0.035} & 0.201{\scriptsize$\pm$0.015} & 0.439{\scriptsize$\pm$0.006} \\
& MedVAE-3D~\citep{varma2025medvae} & \underline{0.613}{\scriptsize$\pm$0.022} & 0.497{\scriptsize$\pm$0.053} & 0.187{\scriptsize$\pm$0.003} & 0.432{\scriptsize$\pm$0.011} \\
 & MedITok~\citep{ma2025meditok} & 0.610{\scriptsize$\pm$0.006} & 0.501{\scriptsize$\pm$0.008} & \underline{0.594}{\scriptsize$\pm$0.031} & \underline{0.568}{\scriptsize$\pm$0.015} \\
 & \cellcolor{gray!20}\methodname{} & \cellcolor{gray!20}\textbf{0.619}{\scriptsize$\pm$0.005} & \cellcolor{gray!20}\textbf{0.570}{\scriptsize$\pm$0.074} & \cellcolor{gray!20}\textbf{0.612}{\scriptsize$\pm$0.006} & \cellcolor{gray!20}\textbf{0.600}{\scriptsize$\pm$0.006} \\
\midrule
\multirow{4}{*}{GREEN ($\uparrow$)}
 & PUMIT~\citep{luo2023pumit} & \underline{0.421}{\scriptsize$\pm$0.017} & 0.204{\scriptsize$\pm$0.010} & 0.124{\scriptsize$\pm$0.019} & 0.250{\scriptsize$\pm$0.009} \\
 & MedVAE-3D~\citep{varma2025medvae} & 0.401{\scriptsize$\pm$0.024} & 0.186{\scriptsize$\pm$0.003} & 0.105{\scriptsize$\pm$0.064} & 0.230{\scriptsize$\pm$0.179} \\
 & MedITok~\citep{ma2025meditok} & 0.412{\scriptsize$\pm$0.025} & \underline{0.256}{\scriptsize$\pm$0.010} & \underline{0.492}{\scriptsize$\pm$0.026} & \underline{0.386}{\scriptsize$\pm$0.013} \\
 & \cellcolor{gray!20}\methodname{} & \cellcolor{gray!20}\textbf{0.425}{\scriptsize$\pm$0.005} & \cellcolor{gray!20}\textbf{0.300}{\scriptsize$\pm$0.078} & \cellcolor{gray!20}\textbf{0.515}{\scriptsize$\pm$0.044} & \cellcolor{gray!20}\textbf{0.413}{\scriptsize$\pm$0.020} \\
\bottomrule
\end{tabularx}
\end{table*}

\paragraph{Visual question answering evaluation.}\label{sec:exp:vqa}
We evaluate each tokenizer as a \textit{frozen} visual backbone within the LLaVA-Med~\cite{li2023llavamed} framework, training only the LLaVA-Med connector and the LLM, and evaluating the resulting models on VQA benchmarks~\cite{ma2025meditok}. We report accuracy assessed via LLM-as-judge grading.
As shown in Table~\ref{tab:vqa}, \methodname achieves the best average accuracy among the competing tokenizers. Notably, \methodname does not use image--text pairs for training, yet still outperforms MedITok, which relies on millions of textual captions for representation learning.

\paragraph{Medical report generation evaluation.}\label{sec:exp:mrg}
Evaluation on generating free-text radiology reports from 3D volumes is similar to VQA, incorporating each frozen tokenizer into the LaMed framework~\cite{bai2024m3d} and training the connector and LLM to generate region-specific findings for three anatomical regions (abdomen, chest, and pelvis).
We report BLEU-2, RaTEScore~\cite{zhao2024ratescore} (entity-level NER-based evaluation), and GREEN~\cite{ostmeier2024green} (clinically aware error detection via an LLM judge).
Table~\ref{tab:mrg} presents the report generation results. \methodname achieves the best scores on all metrics. The improvement is consistent across anatomical regions, with particularly strong gains in the pelvic region, which is the most challenging due to its complex anatomy and longer reference reports. Compared with MedITok, \methodname shows meaningful improvements on clinically aware metrics (GREEN: 0.413 vs.\ 0.386), suggesting that our tokenizer encodes diagnostically relevant information that translates into more accurate clinical language generation. Additional results are provided in \S~\ref{sec:app:eval_setup}.

\subsection{Ablation Studies}\label{sec:exp:ablation}

\begin{figure}[!htbp]
    \centering
    \includegraphics[width=\linewidth]{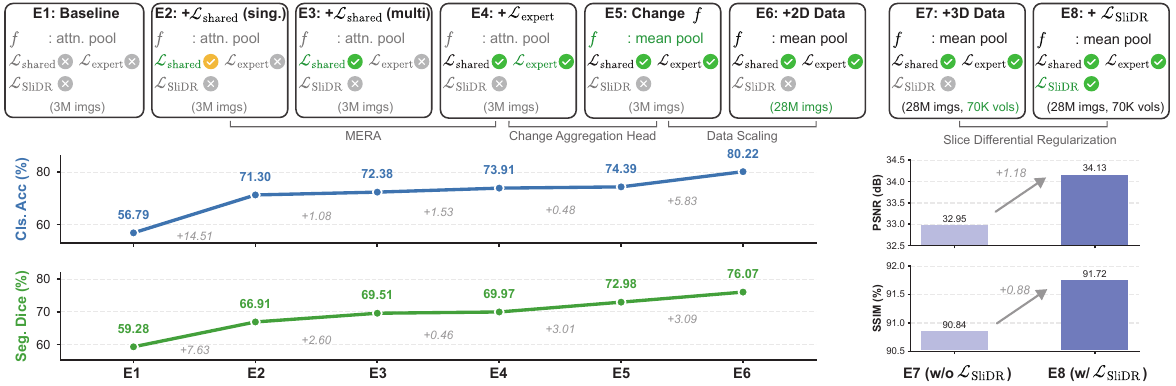}
    \caption{\textbf{Ablation studies.} }
    \label{fig:ablation}
\end{figure}

We ablate the key components of \methodname in Fig.~\ref{fig:ablation}. For efficiency, most components are evaluated on a training subset (3M images) in Stage 1 prior to data scaling.

\paragraph{Effect of MERA components.}
Starting from a baseline without any semantic alignment (E1 in Fig.~\ref{fig:ablation}), shared-space contrastive learning with a single-expert $\mathcal{L}_\text{shared}$ (E2, using MedImageInsight~\cite{codella2024medimageinsight}) yields an initial improvement in classification and segmentation, confirming that semantic supervision is critical for downstream understanding. Extending $\mathcal{L}_\text{shared}$ to multiple experts (E3) further improves both metrics, demonstrating that a panel of modality-specific experts provides more comprehensive semantics. Adding expert-space complementary learning $\mathcal{L}_\text{expert}$ (E4) brings an additional +1.53\% classification gain by preserving modality-specific nuances that shared-space contrastive learning alone may compress away during the shared-space projection.

\paragraph{Aggregation head design.}
Replacing attention pooling~\cite{tschannen2025siglip,lu2025atoken} with mean pooling (E4$\to$E5) yields substantial gains in segmentation. Attention pooling tends to collapse patch-level details into a global summary, whereas mean pooling retains spatially distributed information critical for dense prediction tasks while still achieving competitive classification performance.

\paragraph{Data scaling.}
Scaling the training set from 3M to 28M images (E5$\to$E6) yields consistent improvements, confirming that our framework effectively leverages more diverse training data.

\paragraph{Effect of SliDR.}
Stage~2 (E7) adds 3D data to enable volumetric tokenization. Adding $\mathcal{L}_\text{SliDR}$ (E8) improves 3D reconstruction in terms of both PSNR and SSIM, validating that SliDR specifically strengthens through-plane continuity. Please see \S~\ref{sec:app:abl} for additional analyses.

\FloatBarrier
\Needspace{24\baselineskip}
\section{Conclusion}\label{sec:conclusion}
We present \methodname, a general-purpose medical visual tokenizer that encodes both 2D images and 3D volumes into a shared representational space through a single transformer architecture. Two complementary supervision mechanisms shape this space at different levels: Slice Differential Regularization preserves inter-slice fidelity that per-voxel supervision misses, and Multi-Expert Representation Alignment injects modality-specific clinical semantics from a panel of experts without requiring paired annotations. \methodname achieves state-of-the-art results across reconstruction and six downstream tasks, ranging from pixel-level fidelity to report generation, demonstrating that structural fidelity and semantic richness can be jointly optimized within a single tokenizer. 

We acknowledge several limitations. First, the training data focus on common imaging modalities at present. Other imaging modalities, clinical photos, or illustrations in medical publications remain unexplored. Incorporating these would enhance generalization to non-standard visual inputs encountered in real-world clinical workflows.
Second, while MERA uses frozen expert encoders, jointly fine-tuning selected experts or incorporating new ones could further improve domain coverage when the computational budget allows.
We hope \methodname will serve as a useful building block for scalable multimodal medical AI systems that enable patient-level modeling across heterogeneous imaging evidence and clinical tasks.

\clearpage
\bibliographystyle{unsrtnat}
\bibliography{main}

\clearpage

\appendix

\section*{Appendix}

This appendix provides full experimental details and supplementary results. \S\ref{sec:app_model} presents the tokenizer architecture, including input patchification, attention masking, positional embeddings, and expert encoders. \S\ref{sec:app_train} describes the training settings: \S\ref{sec:app:training_datasets} details the training data construction and lists the 2D and 3D datasets; \S\ref{sec:app:training_setup} specifies optimization, adversarial training, and the two-stage training procedure. \S\ref{sec:app_eval} covers evaluation: \S\ref{sec:app_eval_data} summarizes all evaluation benchmarks grouped by downstream task, \S\ref{sec:app:eval_setup} details the evaluation protocols for each task along with additional quantitative and qualitative results, and \S\ref{sec:app:abl} provides additional ablation studies and analyses. \S\ref{sec:app:broader_impact} discusses broader societal impacts.

\section{Tokenizer Overview}\label{sec:app_model}
\paragraph{Relation to existing tokenizers.}
MedITok~\cite{ma2025meditok} combines reconstruction and semantic alignment for 2D medical images. AToken~\cite{lu2025atoken} provides a unified tokenizer for general-domain images, videos, and 3D assets. SAMTok~\cite{zhou2026samtok} and HiMTok~\cite{wang2025himtok} encode segmentation masks as compact token sequences for multimodal language models. \methodname focuses on a shared representation for medical images and volumes, combining depth-wise reconstruction regularization with semantic supervision from modality-specific experts.

\paragraph{Architecture hyperparameters.}
The encoder and decoder of \methodname each consist of 27 transformer blocks with a hidden dimension of $d=1152$. The encoder is initialized from SigLIP2-SO400M-patch16-NaFlex~\cite{tschannen2025siglip}, while other components are trained from scratch. The patch size is $P_t {\times} P_h {\times} P_w$, where $P_t=4$ is the depth patch size for 3D volumes and $P_h=16, P_w=16$ are the spatial patch sizes for both 2D images and 3D volumes. Rotary position embeddings (RoPE)~\cite{su2024rope} are derived from 3D patch coordinates $(z, x, y)$.
Before the decoder, each encoded $d$-dimensional token embedding is projected to a 96-dimensional output, which is split into mean and log-variance for a diagonal Gaussian posterior, yielding latent codes of dimension $d_z=48$.

\paragraph{Dimension-agnostic input patchification.}
All inputs, regardless of spatial resolution, aspect ratio, or dimensionality, are converted into a common \emph{concatenated patch} format before entering the transformer encoder. Each patch spans $P_t {\times} P_h {\times} P_w = 4 {\times} 16 {\times} 16$ voxels across the depth, height, and width axes, yielding a per-patch feature vector of dimension $P_t \cdot P_h \cdot P_w \cdot C = 3072$ for $C=3$ channels.
For a 2D image of spatial size $H {\times} W$, the input is first zero-padded along the temporal axis to $P_t$ images. Patchification then yields $(H / P_h) \times (W / P_w) = \frac{HW}{256}$ patches (\eg, a $256 {\times} 256$ image produces 256 patches). For a 3D volume of depth $T$ and spatial size $H {\times} W$, patchification directly operates along all three axes, producing $(T / P_t) \times (H / P_h) \times (W / P_w)$ patches (\eg, a 32-slice $256 {\times} 256$ CT volume yields $8 \times 16 \times 16 = 2048$ patches).

All patches from a batch are concatenated into a single flat sequence of shape $[N, 3072]$, where $N$ is the total number of patches across the batch. This eliminates dense padding entirely and achieves $\mathcal{O}(N)$ memory regardless of resolution heterogeneity. Each patch is associated with a coordinate vector $(b, z, x, y)$ encoding the batch index, depth position, and spatial position, together with a list of per-sample sequence lengths that enables $\mathcal{O}(1)$ batch-wise slicing. Because the representation has no global spatial grid, a single batch can freely mix different spatial resolutions (\eg, $256 {\times} 256$ and $512 {\times} 512$), different depths (\eg, 16 and 64 slices), and 2D images with 3D volumes simultaneously.

\paragraph{Attention masking for concatenated sequences.}
Self-attention in the transformer blocks operates over the full concatenated sequence; however, cross-sample information leakage is prevented via block-diagonal attention masks, restricting each token to attend only within the boundaries of its own sample.

\paragraph{Expert encoders for MERA.}
Table~\ref{tab:experts} lists the expert encoders used in MERA. Features from all experts are precomputed offline and stored alongside the training data, so the expert encoders themselves are not loaded during training. Indices 9--10 (3D experts) are included only in Stage~2.

Each input uses the expert assigned to its modality, with a dedicated projector and backprojector. Shared-space alignment brings these representations into one token space, while expert-space alignment retains modality-specific information. Complementary knowledge thus refers to the distinct semantics contributed by experts across modalities.

\begin{table}[h]
    \centering
    \caption{\textbf{Expert encoders used in MERA.}}
    \label{tab:experts}
    \small
    \begin{tabular}{@{}clll@{}}
        \toprule
        Index & Expert Encoder & Domains & $d_j$ \\
        \midrule
        0 & DermLIP~\citep{yan2025derm1m} & Dermoscopy & 768 \\
        1 & RetCLIP~\citep{du2024retclip} & CFP & 768 \\
        2 & Conch~\citep{lu2024conch} & Pathology & 1024 \\
        3 & PeskaVLP~\citep{yuan2024peskavlp} & Endoscopy & 2048 \\
        4--8 & MedImageInsight~\citep{codella2024medimageinsight} & CT, MR, US, X-ray, Other & 2048 \\
        \midrule
        9 & Spectre~\citep{claessens2025spectre} & CT 3D & 2160 \\
        10 & HLIP~\citep{zhao2026hlip} & MR 3D & 2048 \\
        \bottomrule
    \end{tabular}
\end{table}

\section{Training Settings}\label{sec:app_train}

\subsection{Training Datasets}\label{sec:app:training_datasets}

\begin{figure}[p]
    \centering
    \includegraphics[width=\linewidth]{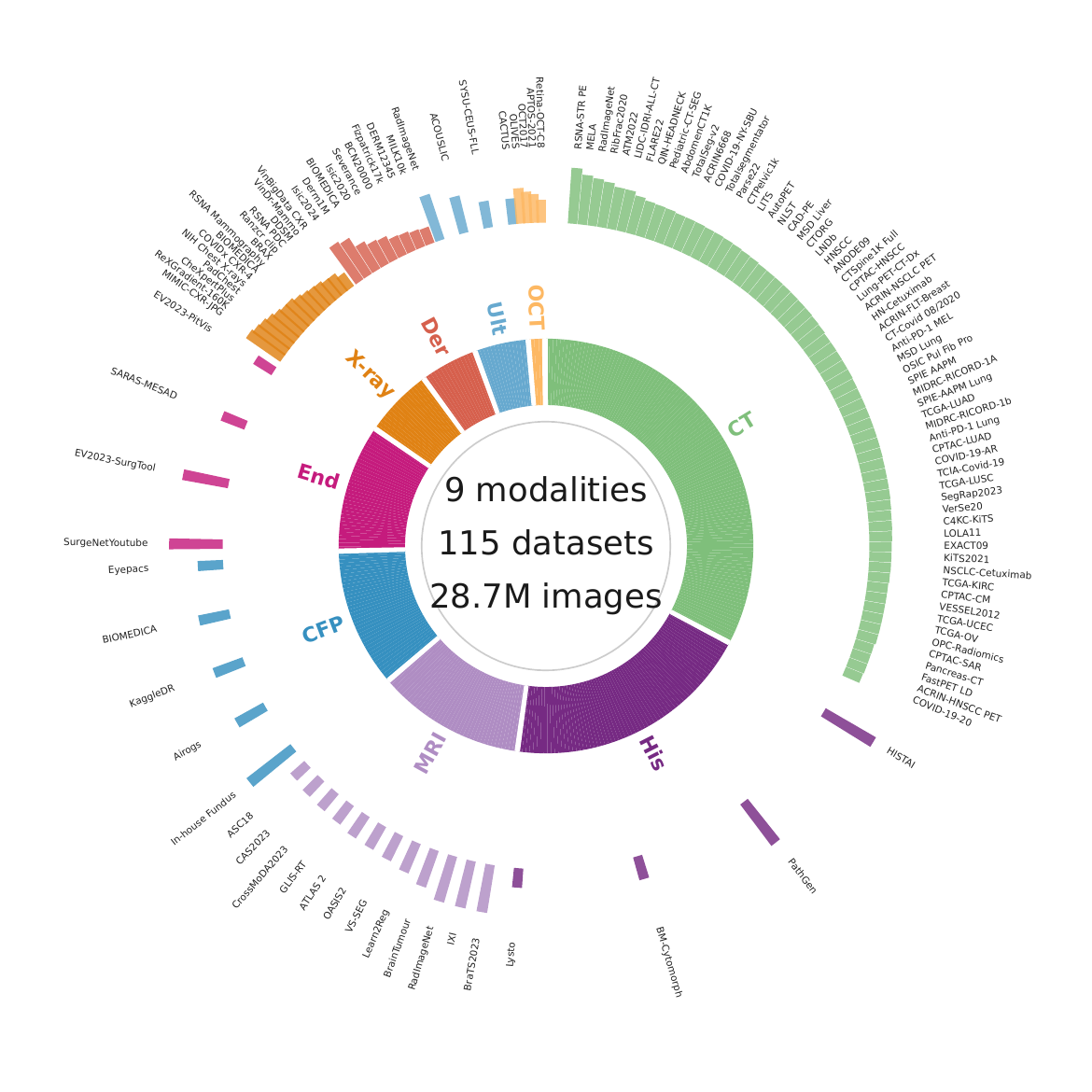}
    \caption{\textbf{Training data distribution across imaging modalities (2D corpus).} 2D corpus comprising 28.7M images across 9 modalities and 115 datasets. The colored mid-ring encodes modality with arc length proportional to its sample count; outer spokes correspond to source datasets, with spoke length scaled by $\log_{10}(\text{count})$. Cross-modality datasets (\eg, RadImageNet, BIOMEDICA, GLIS-RT) are split equally across their constituent modalities. Long dataset names are abbreviated for layout; full names are listed in Table~\ref{tab:datasets_2d}.}
    \label{fig:data_distribution_2d}
\end{figure}

\begin{figure}[p]
    \centering
    \includegraphics[width=\textwidth]{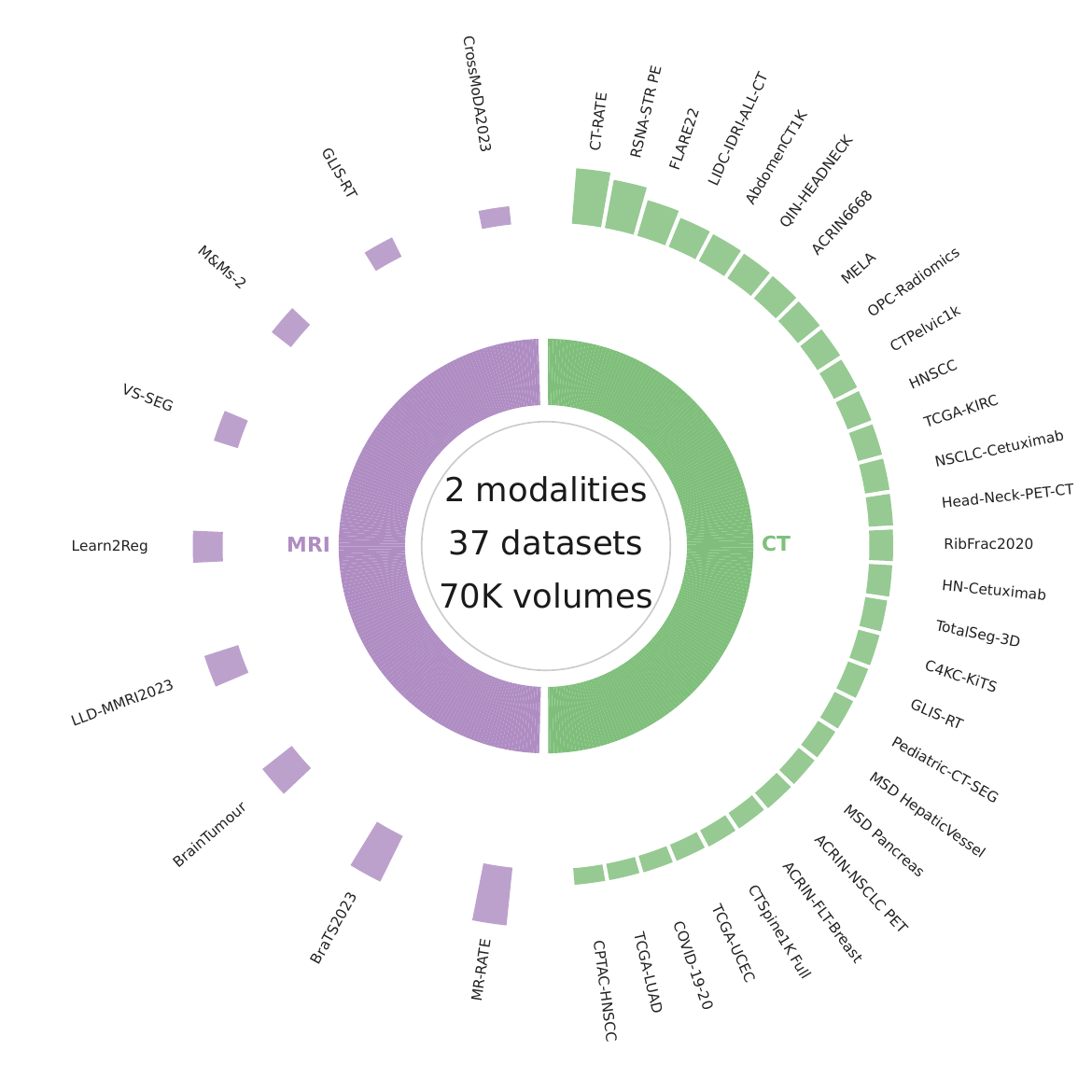}
    \caption{\textbf{Training data distribution across imaging modalities (3D corpus).} 3D corpus comprising 70K volumes across 2 modalities and 37 datasets. The colored mid-ring encodes modality with arc length proportional to its sample count; outer spokes correspond to source datasets, with spoke length scaled by $\log_{10}(\text{count})$. Cross-modality datasets are split equally across their constituent modalities. Long dataset names are abbreviated for layout; full names are listed in Table~\ref{tab:datasets_3d}.}
    \label{fig:data_distribution_3d}
\end{figure}

Figs.~\ref{fig:data_distribution_2d} and~\ref{fig:data_distribution_3d} provide a visual overview of the training data composition; Tables~\ref{tab:datasets_2d} and~\ref{tab:datasets_3d} list the 2D and 3D training datasets, respectively. In total, we collect 28,769,502 images and 70,144 volumes spanning 9 imaging modalities.

We curate the training set from publicly available medical imaging repositories. 
Before quality filtering, we impose minimum dataset size thresholds: only 2D datasets containing at least 9{,}000 images and 3D datasets containing at least 180 volumes are included. This criterion serves two purposes. First, it prevents commonly used small-scale benchmark datasets from entering the training pool, thereby preserving their integrity for downstream evaluation. Second, it avoids highly fragmented data sources that contribute minimal training signal relative to the overhead of integration.

Following MedITok~\citep{ma2025meditok}, we apply a series of automated quality filters to remove low-quality or clinically irrelevant 2D images, including checks on dynamic range, minimum resolution, pixel-value diversity, and information content, as well as manual screening to exclude non-clinical illustrations such as tables and plots.

\subsection{Implementation Details}\label{sec:app:training_setup}

As stated in the main text (Eq.~\ref{eq:total_loss}), the total training objective is $\mathcal{L} = \mathcal{L}_\text{R} + \lambda_\text{SliDR}\mathcal{L}_\text{SliDR} + \lambda_\text{KL}\mathcal{L}_\text{KL} + \lambda_\text{MERA}\mathcal{L}_\text{MERA}$. Below we detail the loss components, the two-stage training procedure, and the optimization settings.

\paragraph{Loss functions.}
The reconstruction loss is defined as:
\begin{equation}
    \mathcal{L}_\text{R} = \ell_1 + \ell_2 + \mathcal{L}_\text{LPIPS} + w_\text{adv} \mathcal{L}_\text{adv},
\end{equation}
where $\ell_1$ and $\ell_2$ denote mean absolute error and mean squared error, respectively; $\mathcal{L}_\text{LPIPS}$~\cite{zhang2018unreasonable} is the perceptual loss; and $\mathcal{L}_\text{adv}$ is the adversarial loss with an adaptive weight $w_\text{adv}$ that balances reconstruction and adversarial gradients on the decoder's last layer, following VQGAN~\citep{esser2021taming}.

For the KL regularization, patch tokens $\vct{h} \in \mathbb{R}^{N \times d}$ are linearly projected to a mean $\vct{\mu} \in \mathbb{R}^{N\times d_z}$ and a log-variance $\log \vct{\sigma}^2 \in \mathbb{R}^{N\times d_z}$, where $d_z=48$. The latent code is sampled via reparameterization as $\vct{z} = \vct{\mu} + \vct{\sigma} \odot \vct{\epsilon}$, $\vct{\epsilon} \sim \mathcal{N}(\mathbf{0}, \mathbf{I})$. The KL term penalizes the deviation of the diagonal Gaussian posterior from the standard normal prior:
\begin{equation}
    \mathcal{L}_\text{KL} = \frac{1}{2} \sum_{n,k} \bigl( \vct{\mu}_{n,k}^2 + \vct{\sigma}_{n,k}^2 - 1 - \log \vct{\sigma}_{n,k}^2 \bigr),
\end{equation}
weighted by $\lambda_\text{KL} = 10^{-6}$ following Rombach \etal~\cite{rombach2022high}. We set $\lambda_\text{MERA} = 1$ and $\lambda_\text{SliDR}=0.1$ by default; ablation studies on these coefficients can be found in Sec.~\ref{sec:app:abl}.

\paragraph{Two-stage training.}
In Stage~1, \methodname is trained exclusively on 2D images with $\mathcal{L}_\text{R}$, $\mathcal{L}_\text{KL}$, and $\mathcal{L}_\text{MERA}$ ($\lambda_\text{SliDR}=0$). Only 2D expert projectors are active. A diversity sampler ensures that each mini-batch contains at least two samples from every available modality group, with the remaining slots filled from the global pool.

In Stage~2, we initialize from the Stage~1 checkpoint, introduce 3D expert projectors (indices 9--10), and activate $\mathcal{L}_\text{SliDR}$ with $\lambda_\text{SliDR}=0.1$. Training alternates between 2D and 3D batches at every iteration: even iterations process a 2D batch (effective batch size 600) and odd iterations process a 3D batch (effective batch size 48). The 3D dataloader cycles indefinitely to maintain continuous 3D supervision regardless of the smaller 3D dataset size. Within each 3D batch, a balanced sampler maintains an approximately equal ratio of CT and MR volumes.

\paragraph{Optimization.}
We use AdamW~\cite{loshchilov2018adamw} for both the tokenizer and discriminator. The tokenizer uses $(\beta_1, \beta_2) = (0.9, 0.95)$ with a cosine annealing schedule: in Stage~1, the learning rate decays from $10^{-4}$ to $10^{-5}$ over ${\sim}$100K steps; in Stage~2, from $2 \times 10^{-5}$ to $2 \times 10^{-6}$ over ${\sim}$100K steps. The discriminator uses $(\beta_1, \beta_2) = (0.5, 0.9)$. All training uses BFloat16 mixed precision. During training, images are of mixed resolutions: 256, 384, and 512.

\paragraph{Compute resources.}
Full-scale data training is conducted on 16$\times$ NVIDIA H200 GPUs (141 GB); peak GPU memory usage is 132.54\,GB per device. Each compute node is an x86\_64 machine with dual Intel Xeon Platinum 8558 processors (2 sockets, 48 physical cores per socket, 2 threads per core). Stage~1 takes approximately 15.75 hours and Stage~2 takes approximately 43.5 hours. The full research project required more compute than reported here due to preliminary and failed experiments during hyperparameter exploration and architecture design.

\paragraph{License.}
The code and model weights will be released under the Apache 2.0 license.

\section{Evaluation}\label{sec:app_eval}
\subsection{Evaluation Datasets}\label{sec:app_eval_data}
Table~\ref{tab:eval_datasets} summarizes all evaluation datasets used in our experiments, grouped by downstream task. We verified that all evaluation datasets listed are excluded from the training corpus.

\begin{table*}[h]
\centering
\caption{\textbf{Evaluation datasets grouped by downstream task.} Rec: reconstruction, Cls: classification, Ret: retrieval, Seg: segmentation, Syn: image synthesis, VQA: visual question answering, MRG: medical report generation. \#Cls denotes the number of target classes. Datasets appearing in multiple task groups are listed separately.}
\label{tab:eval_datasets}
\footnotesize
\setlength{\tabcolsep}{5pt}
\begin{tabular}{@{}llrrl@{}}
\toprule
Dataset & Modality & \#Samples & \#Cls & Task(s) \\
\midrule
\multicolumn{5}{l}{\textit{2D Perception -- Reconstruction / Classification / Retrieval}} \\
ChestCTScan~\cite{syed2024medicalimagining} & CT & 1,000 & 4 & Rec, Cls, Ret \\
JSIEC~\cite{cen2021automatic} & CFP & 1,000 & 39 & Rec, Cls, Ret \\
HAM10K~\cite{tschandl2018ham10000} & Dermoscopy & 10,015 & 7 & Rec, Cls, Ret \\
PolypDiag~\cite{tian2022contrastive} & Endoscopy & 2,202 & 2 & Rec, Cls, Ret \\
BRISC~\cite{fateh2025brisc} & MRI & 5,023 & 4 & Rec, Cls, Ret \\
OCTDL~\cite{kulyabin2024octdl} & OCT & 3,236 & 7 & Rec, Cls, Ret \\
LungHist700~\cite{diosdado2024lunghist700} & Pathology & 691 & 3 & Rec, Cls, Ret \\
BUSBRA~\cite{gomez2024bus} & Ultrasound & 1,875 & 4 & Rec, Cls, Ret \\
Kermany~\cite{kermany2018identifying} & X-ray & 5,856 & 2 & Rec, Cls, Ret \\
\midrule
\multicolumn{5}{l}{\textit{2D Perception -- Segmentation}} \\
MaplesDR~\cite{lepetit2024maples} & CFP & 192 & 3 & Seg \\
MosMedDataPlus~\cite{morozov2020mosmeddata} & CT & 2,729 & 2 & Seg \\
HAM10K~\cite{tschandl2018ham10000} & Dermoscopy & 10,015 & 2 & Seg \\
PolypDB~\cite{jha2024polypdb} & Endoscopy & 3,843 & 2 & Seg \\
BRISC~\cite{fateh2025brisc} & MRI & 4,746 & 2 & Seg \\
PanNuke~\cite{gamper2019pannuke} & Pathology & 7,901 & 6 & Seg \\
BUSBRA~\cite{gomez2024bus} & Ultrasound & 1,875 & 2 & Seg \\
VinDr-RibCXR~\cite{nguyen2021vindr} & X-ray & 245 & 2 & Seg \\
\midrule
\multicolumn{5}{l}{\textit{3D Volumetric Reconstruction}} \\
AMOS-MM~\cite{ji2022amos} & CT & 100 & -- & 3D Rec \\
MRISegmenter~\cite{zhuang2025mrisegmenter} & MRI & 240 & -- & 3D Rec \\
\midrule
\multicolumn{5}{l}{\textit{Image Synthesis}} \\
GMAI-VL-5.5M~\cite{li2026gmai} & Multi-modality & 70,000 & -- & Syn \\
\midrule
\multicolumn{5}{l}{\textit{Visual Question Answering (test benchmarks)}} \\
Fundus-MMBench~\cite{liu2025constructing} & CFP & 620 & -- & VQA \\
GMAI-MMBench~\cite{chen2024gmai} & Multi-modality & 4,550 & -- & VQA \\
Kvasir-VQA-x1~\cite{gautam2025kvasir} & Endoscopy & 15,955 & -- & VQA \\
MM-Skin~\cite{zeng2025mm} & Dermoscopy & 5,481 & -- & VQA \\
PathVQA~\cite{he2020pathvqa} & Pathology & 6,719 & -- & VQA \\
SLAKE~\cite{liu2021slake} & Radiography & 2,094 & -- & VQA \\
\midrule
\multicolumn{5}{l}{\textit{Medical Report Generation}} \\
AMOS-MM~\cite{ji2022amos} & CT & 1,687 & -- & MRG \\
\bottomrule
\end{tabular}
\end{table*}

\begin{figure}[ht]
    \centering
    \includegraphics[width=\linewidth]{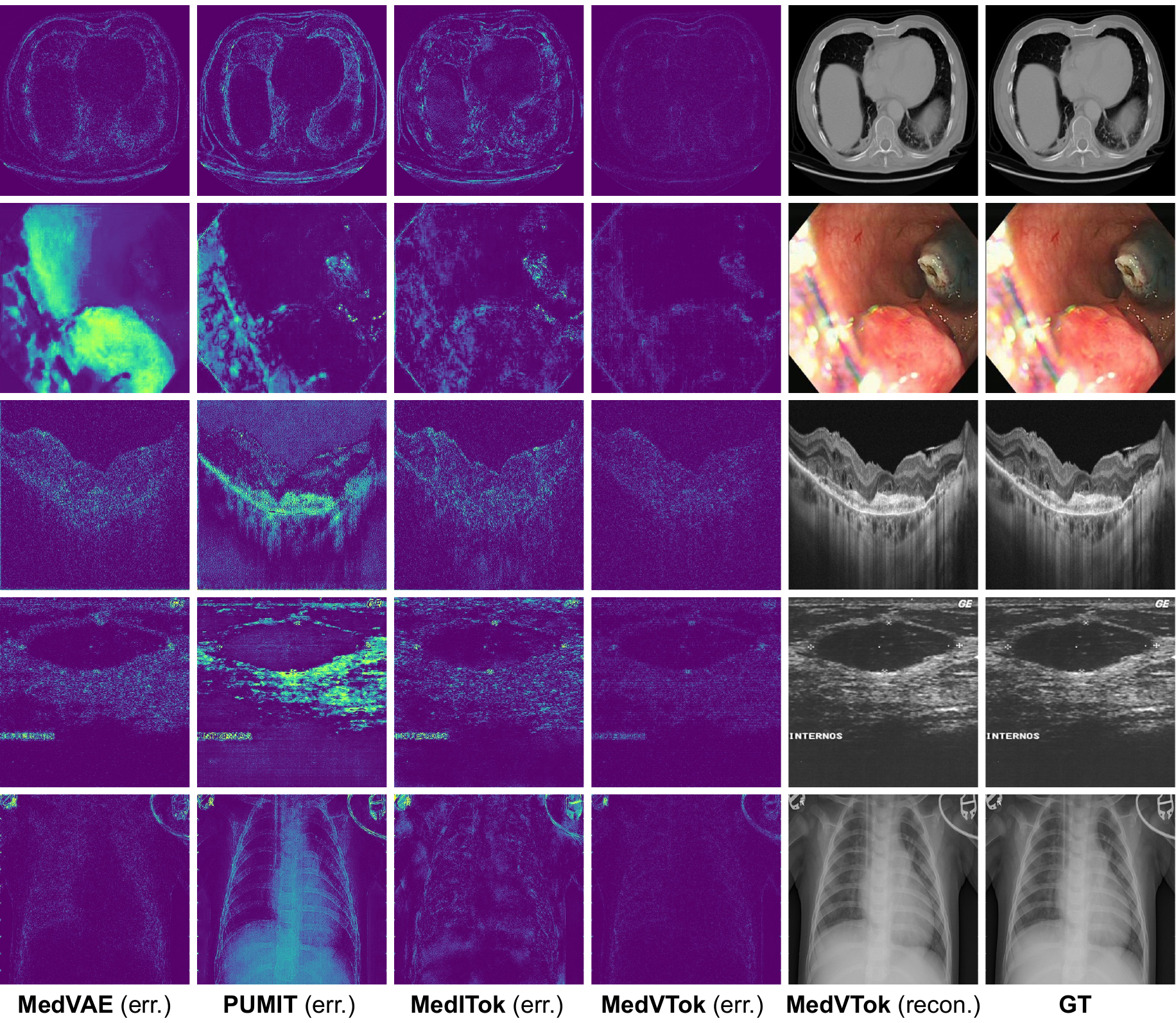}
    \caption{\textbf{Additional reconstruction visualization.}}
    \label{fig:result_recon_2d_supp}
\end{figure}

\begin{figure}[ht]
    \centering
    \includegraphics[width=\linewidth]{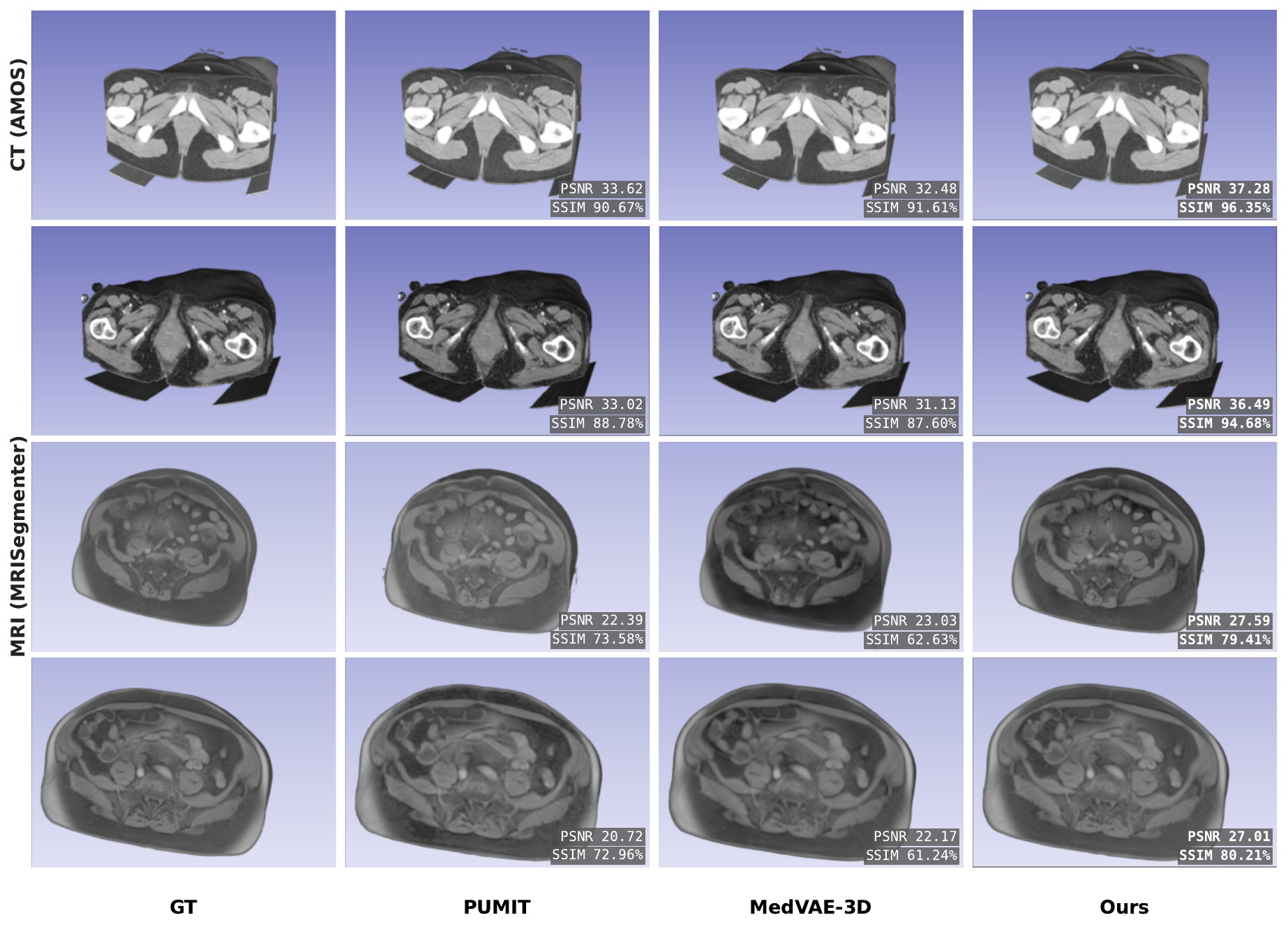}
    \caption{\textbf{3D Reconstruction visualization.}}
    \label{fig:result_3d_recon}
\end{figure}

\subsection{Evaluation Setup and Results}\label{sec:app:eval_setup}

\paragraph{Token fidelity.}
We use autoencoding reconstruction as a proxy task to evaluate token fidelity: each input is encoded and decoded by the frozen tokenizer, and we measure how faithfully the tokens preserve the original signal.
For 2D images, evaluation is performed at $512 {\times} 512$ resolution. We report PSNR and SSIM between the input and reconstructed images.
For 3D volumes, we evaluate on CT and MRI using two datasets. CT intensities are clipped to $[-1000, 1000]$~HU; MRI intensities are clipped at the 0.5th and 99.5th percentiles per volume. PSNR and SSIM are computed at the volume level.
Only models with native 3D support (PUMIT, MedVAE, AToken, \methodname) are included in the 3D evaluation. 

Additional visualizations of reconstructions on 2D and 3D modalities are presented in Fig.~\ref{fig:result_recon_2d_supp} and Fig.~\ref{fig:result_3d_recon}, respectively.

\paragraph{Classification.}
We adopt a $k$-nearest neighbor ($k$-NN) protocol on frozen features extracted from the CLS token (or equivalent global representation) of each tokenizer. For tokenizers that do not output such a representation (\eg, SDVAE), we concatenate all features along the channel dimension. The feature vectors are $L_2$-normalized, and classification is performed via softmax-weighted voting over the $k$ nearest neighbors using cosine similarity with a temperature of $T=0.07$. We sweep $k \in \{10, 20, 100, 200\}$ on the validation set and report the best accuracy for each model. Image resolution is controlled at $256 {\times} 256$ when possible.

\paragraph{Retrieval.}
We evaluate image-to-image retrieval using the same frozen features and benchmarks as classification. The training split of each dataset serves as the gallery, and the test split provides the queries. For each query, we compute the cosine similarity between its $L_2$-normalized feature vector and all gallery features, rank the gallery by descending similarity, and report Recall@1 (the fraction of queries whose top-retrieved image shares the same label). Image resolution is controlled at $256 {\times} 256$ when possible, consistent with the classification setup.

\paragraph{Segmentation.}
We evaluate the quality of frozen patch tokens for dense prediction by training a lightweight linear segmentation head on top of each tokenizer's output, following the protocol of Sim{\'e}oni \etal~\cite{simeoni2025dinov3}. The segmentation head consists of a batch normalization layer followed by a $1 {\times} 1$ convolution. For each tokenizer, we train separate heads under identical settings and select the best-performing one by validation Dice score.
Training runs for 50 epochs with AdamW (no weight decay), a batch size of 32, cross-entropy loss, and learning rates searched over $\{10^{-2}, 10^{-3}, 10^{-4}\}$. Image resolution is controlled at $256 {\times} 256$ when possible. We report the overall Dice score over segmentation classes. Table~\ref{tab:seg_hd95} further reports Hausdorff Distance at the 95th percentile (HD95), where \methodname still leads in terms of average performance across modalities.

\begin{table*}[t]
\centering
\caption{\textbf{Additional results on image segmentation evaluation.} Method type: $^{*}$semantic encoder; $^{\ddagger}$unified tokenizer.}
\label{tab:seg_hd95}
\resizebox{\textwidth}{!}{
\begin{tabular}{llccccccccc}
\toprule
Metrics & Method & CFP & CT & Derm. & Endo. & MRI & Path. & US & X-ray & Avg. \\
\midrule
 \multirow{13}{*}{\shortstack{Dice ($\uparrow$)}} 
  & SDVAE~\citep{rombach2022high} & 73.03 & 14.88 & 62.37 & 24.67 & 12.69 & 49.11 & 84.19 & 35.56 & 44.56 \\
  & TokenFlow$^{\ddagger}$~\citep{qu2024tokenflow} & 70.39 & 19.05 & 72.62 & 40.53 & 30.49 & 52.34 & 91.33 & 49.52 & 53.28 \\
  & MingTok$^{\ddagger}$~\citep{huang2025mingunivision} & 85.36 & 27.03 & 83.91 & 57.32 & 52.54 & 67.68 & 92.91 & 54.71 & 65.18 \\
  & UniTok$^{\ddagger}$~\citep{ma2025unitok} & 60.93 & 33.13 & 86.25 & 67.25 & 54.90 & 72.51 & 93.35 & 57.60 & 65.74 \\
  & AToken$^{\ddagger}$~\citep{lu2025atoken} & 97.09 & 37.33 & 87.45 & \underline{69.88} & 63.57 & 74.30 & 94.90 & 59.69 & 73.03 \\
  \cmidrule{2-11}
  & BiomedCLIP$^{*}$~\citep{zhang2023biomedclip} & 50.22 & 31.83 & 87.48 & 60.64 & 57.95 & 73.65 & 95.41 & 59.83 & 64.63 \\
  & MedSigLIP$^{*}$~\citep{sellergren2025medgemma} & \underline{97.28} & \underline{40.79} & \underline{87.76} & 69.81 & \underline{67.32} & \underline{76.20} & 95.33 & \textbf{62.48} & \underline{74.62} \\
  & MedImageInsight$^{*}$~\citep{codella2024medimageinsight} & 92.34 & 25.60 & 85.45 & 66.96 & 55.81 & 42.93 & \underline{95.80} & 50.16 & 64.38 \\
  & PUMIT~\citep{luo2023pumit} & 40.05 & 25.53 & 77.94 & 37.32 & 36.69 & 66.00 & 90.73 & 54.82 & 53.63 \\
  & MedVAE~\citep{varma2025medvae} & 61.74 & 12.57 & 57.59 & 19.73 & 10.13 & 36.40 & 81.90 & 30.81 & 38.86 \\
  & MedITok$^{\ddagger}$~\citep{ma2025meditok} & 96.70 & 40.52 & 80.28 & 66.07 & 62.42 & 75.80 & 94.65 & 59.47 & 71.99 \\
  & \cellcolor{gray!20}\methodname{}$^{\ddagger}$ & \cellcolor{gray!20}\textbf{97.44} & \cellcolor{gray!20}\textbf{42.91} & \cellcolor{gray!20}\textbf{90.25} & \cellcolor{gray!20}\textbf{72.52} & \cellcolor{gray!20}\textbf{69.62} & \cellcolor{gray!20}\textbf{77.74} & \cellcolor{gray!20}\textbf{96.91} & \cellcolor{gray!20}\underline{61.13} & \cellcolor{gray!20}\textbf{76.07} \\
\midrule
 \multirow{13}{*}{\shortstack{HD95 ($\downarrow$)}} 
  & SDVAE~\citep{rombach2022high} & 109.69 & 218.98 & 56.30 & 124.25 & 343.33 & 241.99 & 172.77 & 46.39 & 164.21 \\
  & TokenFlow$^{\ddagger}$~\citep{qu2024tokenflow} & 57.43 & 211.68 & 56.90 & 73.33 & 151.20 & 278.48 & 45.72 & 23.82 & 112.32 \\
  & MingTok$^{\ddagger}$~\citep{huang2025mingunivision} & 115.80 & 180.54 & 40.70 & 109.14 & 76.67 & 230.27 & 51.49 & 34.18 & 104.85 \\
  & UniTok$^{\ddagger}$~\citep{ma2025unitok} & 104.54 & 123.44 & 18.10 & 51.08 & 66.77 & 162.47 & 22.59 & 13.85 & 70.35 \\
  & AToken$^{\ddagger}$~\citep{lu2025atoken} & \underline{22.99} & 119.80 & 16.05 & 47.32 & 29.93 & 160.72 & 16.64 & \textbf{12.63} & 53.26 \\
  \cmidrule{2-11}
  & BiomedCLIP$^{*}$~\citep{zhang2023biomedclip} & 84.43 & 178.88 & \underline{12.62} & 50.33 & 31.22 & 145.42 & 13.75 & \underline{12.97} & 66.20 \\
  & MedSigLIP$^{*}$~\citep{sellergren2025medgemma} & 38.47 & \underline{112.54} & 13.94 & \textbf{35.48} & \textbf{21.55} & \underline{138.56} & 16.82 & 21.45 & \underline{49.85} \\
  & MedImageInsight$^{*}$~\citep{codella2024medimageinsight} & 31.83 & 248.33 & 17.16 & 58.84 & 54.94 & 174.84 & \underline{12.50} & 27.52 & 78.24 \\
  & PUMIT~\citep{luo2023pumit} & 63.10 & 151.79 & 30.78 & 76.98 & 78.13 & 214.75 & 35.29 & 19.50 & 83.79 \\
  & MedVAE~\citep{varma2025medvae} & 326.48 & 241.19 & 69.08 & 102.19 & 164.70 & 343.04 & 481.75 & 47.13 & 221.94 \\
  & MedITok$^{\ddagger}$~\citep{ma2025meditok} & 44.53 & 138.61 & 37.06 & 46.60 & 32.97 & 140.30 & 18.31 & 13.85 & 59.03 \\
  & \cellcolor{gray!20}\methodname{} & \cellcolor{gray!20}\textbf{19.12} & \cellcolor{gray!20}\textbf{105.06} & \cellcolor{gray!20}\textbf{11.67} & \cellcolor{gray!20}\underline{46.31} & \cellcolor{gray!20}\underline{24.76} & \cellcolor{gray!20}\textbf{121.30} & \cellcolor{gray!20}\textbf{12.43} & \cellcolor{gray!20}15.21 & \cellcolor{gray!20}\textbf{44.48} \\
\bottomrule
\end{tabular}
}
\end{table*}

\begin{figure}[ht]
    \centering
    \includegraphics[width=\linewidth]{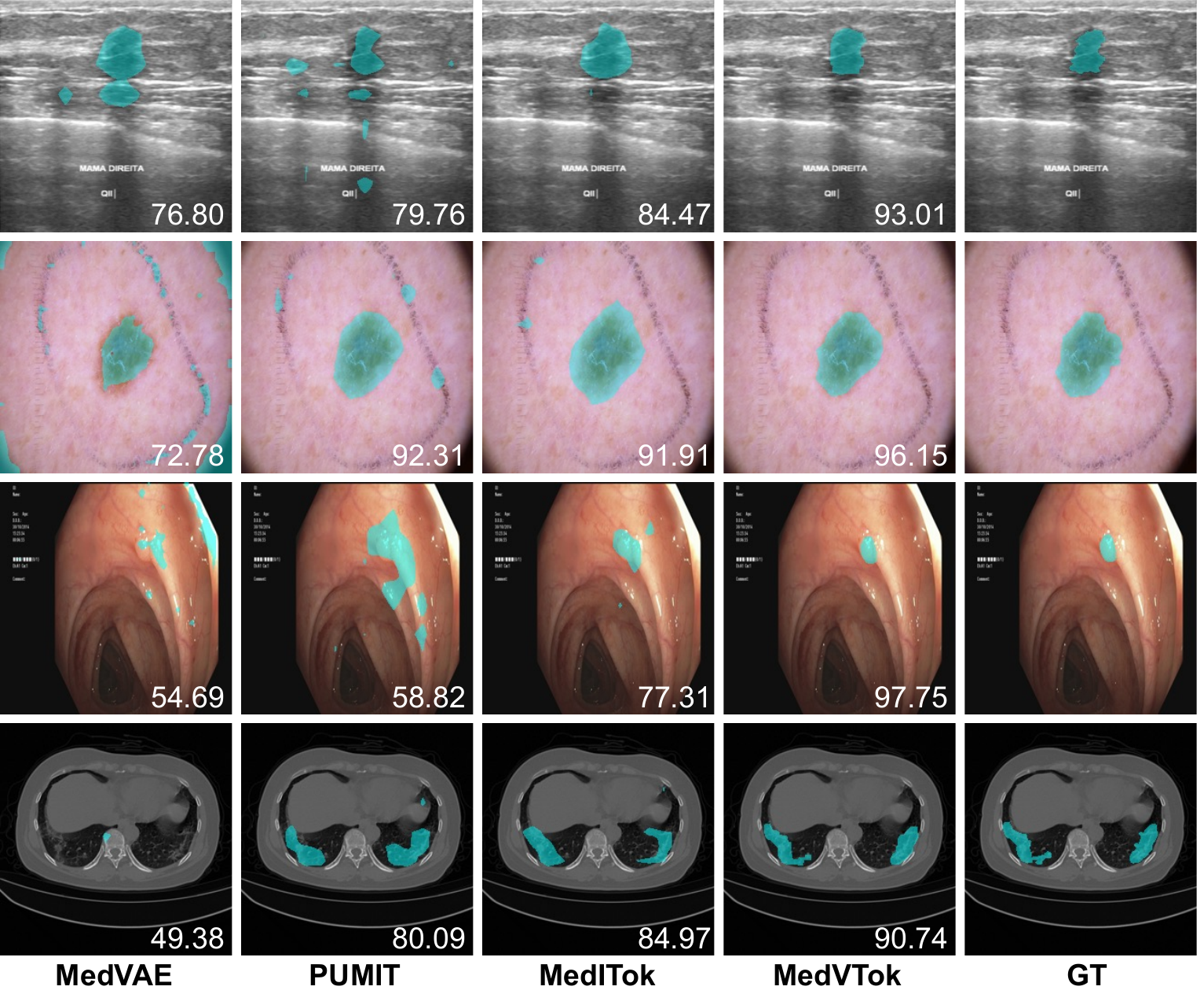}
    \caption{\textbf{Qualitative comparison on the image segmentation task.} Segmentation powered by \methodname achieves the best performance, while other tokenizers lead to false segmentation in the background. In each image, the mask is highlighted in cyan; the Dice score is shown in the lower right corner.}
    \label{fig:result_seg}
\end{figure}

\paragraph{Comparison with modality experts.}
Tables~\ref{tab:expert_cls} and~\ref{tab:expert_seg} compare \methodname with the experts used in MERA under the classification and segmentation protocols above. \methodname exceeds MedImageInsight on seven of nine classification benchmarks and all eight segmentation benchmarks. Among the four specialized experts, it exceeds PeskaVLP in classification and RetCLIP, DermLIP, and PeskaVLP in segmentation. Conch retains higher scores on pathology.

\begin{table}[!htbp]
\centering
\caption{\textbf{Classification accuracy (\%) compared with MERA experts.} Specialized experts are evaluated on their respective modalities. Values are means $\pm$ 95\% confidence intervals.}
\label{tab:expert_cls}
\scriptsize
\setlength{\tabcolsep}{1.5pt}
\begin{tabularx}{\linewidth}{@{}l*{9}{>{\centering\arraybackslash}X}@{}}
\toprule
Model & CFP & CT & Derm. & Endo. & MRI & OCT & Path. & US & X-ray \\
\midrule
RetCLIP & \mbox{\textbf{91.95}{\scriptsize$\pm$1.05}} & -- & -- & -- & -- & -- & -- & -- & -- \\
DermLIP & -- & -- & \mbox{79.72{\scriptsize$\pm$0.66}} & -- & -- & -- & -- & -- & -- \\
PeskaVLP & -- & -- & -- & \mbox{48.64{\scriptsize$\pm$2.29}} & -- & -- & -- & -- & -- \\
Conch & -- & -- & -- & -- & -- & -- & \mbox{\textbf{93.08}{\scriptsize$\pm$2.44}} & -- & -- \\
MedImageInsight & \mbox{75.64{\scriptsize$\pm$2.12}} & \mbox{75.32{\scriptsize$\pm$2.83}} & \mbox{\textbf{85.87}{\scriptsize$\pm$0.38}} & \mbox{44.65{\scriptsize$\pm$0.74}} & \mbox{\textbf{95.47}{\scriptsize$\pm$0.29}} & \mbox{90.79{\scriptsize$\pm$1.18}} & \mbox{71.79{\scriptsize$\pm$2.92}} & \mbox{50.18{\scriptsize$\pm$2.67}} & \mbox{96.21{\scriptsize$\pm$0.83}} \\
\methodname{} & \mbox{90.54{\scriptsize$\pm$1.05}} & \mbox{\textbf{87.03}{\scriptsize$\pm$1.64}} & \mbox{75.27{\scriptsize$\pm$0.99}} & \mbox{\textbf{51.00}{\scriptsize$\pm$1.85}} & \mbox{95.03{\scriptsize$\pm$0.52}} & \mbox{\textbf{94.00}{\scriptsize$\pm$1.18}} & \mbox{78.97{\scriptsize$\pm$1.65}} & \mbox{\textbf{52.92}{\scriptsize$\pm$1.68}} & \mbox{\textbf{97.22}{\scriptsize$\pm$0.61}} \\
\bottomrule
\end{tabularx}
\end{table}

\begin{table}[!htbp]
\centering
\caption{\textbf{Segmentation Dice (\%) compared with MERA experts.} Specialized experts are evaluated on their respective modalities. Values are means $\pm$ 95\% confidence intervals.}
\label{tab:expert_seg}
\footnotesize
\setlength{\tabcolsep}{1.5pt}
\begin{tabularx}{\linewidth}{@{}l*{8}{>{\centering\arraybackslash}X}@{}}
\toprule
Model & CFP & CT & Derm. & Endo. & MRI & Path. & US & X-ray \\
\midrule
RetCLIP & \mbox{92.46{\scriptsize$\pm$1.69}} & -- & -- & -- & -- & -- & -- & -- \\
DermLIP & -- & -- & \mbox{89.05{\scriptsize$\pm$1.86}} & -- & -- & -- & -- & -- \\
PeskaVLP & -- & -- & -- & \mbox{65.06{\scriptsize$\pm$1.51}} & -- & -- & -- & -- \\
Conch & -- & -- & -- & -- & -- & \mbox{\textbf{83.77}{\scriptsize$\pm$1.39}} & -- & -- \\
MedImageInsight & \mbox{92.34{\scriptsize$\pm$1.39}} & \mbox{25.60{\scriptsize$\pm$1.35}} & \mbox{85.45{\scriptsize$\pm$2.10}} & \mbox{66.96{\scriptsize$\pm$1.32}} & \mbox{55.81{\scriptsize$\pm$1.31}} & \mbox{42.93{\scriptsize$\pm$1.86}} & \mbox{95.80{\scriptsize$\pm$2.07}} & \mbox{50.16{\scriptsize$\pm$1.48}} \\
\methodname{} & \mbox{\textbf{97.44}{\scriptsize$\pm$1.59}} & \mbox{\textbf{42.91}{\scriptsize$\pm$1.80}} & \mbox{\textbf{90.25}{\scriptsize$\pm$0.77}} & \mbox{\textbf{72.52}{\scriptsize$\pm$1.15}} & \mbox{\textbf{69.62}{\scriptsize$\pm$1.07}} & \mbox{77.74{\scriptsize$\pm$1.56}} & \mbox{\textbf{96.91}{\scriptsize$\pm$1.80}} & \mbox{\textbf{61.13}{\scriptsize$\pm$1.17}} \\
\bottomrule
\end{tabularx}
\end{table}

\paragraph{3D segmentation.}
We evaluate frozen tokenizers on AMOS-MM CT~\cite{ji2022amos} and MRISegmenter MRI~\cite{zhuang2025mrisegmenter}, both excluded from tokenizer pretraining. For each tokenizer, we train a lightweight 3D convolutional head on volumetric features to predict segmentation masks. Table~\ref{tab:seg3d} reports overall and per-organ Dice. \methodname achieves the highest overall Dice on both datasets.

\begin{table}[!htbp]
\centering
\caption{\textbf{3D segmentation Dice (\%).} Each tokenizer is frozen while a lightweight 3D convolutional head is trained. Values include reported 95\% confidence intervals.}
\label{tab:seg3d}
\footnotesize
\setlength{\tabcolsep}{1.5pt}
\begin{tabularx}{\linewidth}{@{}l*{8}{>{\centering\arraybackslash}X}@{}}
\toprule
 & \multicolumn{4}{c}{AMOS-MM CT} & \multicolumn{4}{c}{MRISegmenter MRI} \\
\cmidrule(lr){2-5}\cmidrule(lr){6-9}
Organ & PUMIT & \shortstack{MedVAE\\3D} & AToken & \methodname{} & PUMIT & \shortstack{MedVAE\\3D} & AToken & \methodname{} \\
\midrule
Overall & 28.67 & 29.71 & 36.89 & \textbf{50.13} & \mbox{44.57{\scriptsize$\pm$0.76}} & \mbox{43.70{\scriptsize$\pm$0.80}} & \mbox{54.63{\scriptsize$\pm$0.68}} & \mbox{\textbf{57.20}{\scriptsize$\pm$0.76}} \\
Spleen & \mbox{63.68{\scriptsize$\pm$1.58}} & \mbox{42.94{\scriptsize$\pm$2.64}} & \mbox{55.76{\scriptsize$\pm$2.46}} & \mbox{\textbf{70.61}{\scriptsize$\pm$2.30}} & \mbox{40.83{\scriptsize$\pm$2.12}} & \mbox{40.68{\scriptsize$\pm$2.14}} & \mbox{63.28{\scriptsize$\pm$2.71}} & \mbox{\textbf{70.60}{\scriptsize$\pm$2.85}} \\
Kidney & \mbox{48.94{\scriptsize$\pm$1.96}} & \mbox{45.95{\scriptsize$\pm$2.36}} & \mbox{59.56{\scriptsize$\pm$2.44}} & \mbox{\textbf{74.93}{\scriptsize$\pm$1.74}} & \mbox{81.60{\scriptsize$\pm$1.29}} & \mbox{77.46{\scriptsize$\pm$1.77}} & \mbox{83.96{\scriptsize$\pm$0.67}} & \mbox{\textbf{86.48}{\scriptsize$\pm$0.95}} \\
Liver & \mbox{76.63{\scriptsize$\pm$2.68}} & \mbox{73.73{\scriptsize$\pm$2.23}} & \mbox{80.50{\scriptsize$\pm$1.72}} & \mbox{\textbf{86.31}{\scriptsize$\pm$1.00}} & \mbox{58.11{\scriptsize$\pm$1.32}} & \mbox{59.11{\scriptsize$\pm$1.23}} & \mbox{85.71{\scriptsize$\pm$0.94}} & \mbox{\textbf{85.83}{\scriptsize$\pm$0.70}} \\
Pancreas & \mbox{13.42{\scriptsize$\pm$1.99}} & \mbox{23.51{\scriptsize$\pm$2.60}} & \mbox{31.64{\scriptsize$\pm$2.72}} & \mbox{\textbf{41.90}{\scriptsize$\pm$2.36}} & \mbox{49.43{\scriptsize$\pm$2.22}} & \mbox{43.98{\scriptsize$\pm$2.08}} & \mbox{49.23{\scriptsize$\pm$1.92}} & \mbox{\textbf{49.49}{\scriptsize$\pm$2.71}} \\
Adrenal & \mbox{16.50{\scriptsize$\pm$1.30}} & \mbox{12.74{\scriptsize$\pm$1.10}} & \mbox{20.44{\scriptsize$\pm$1.47}} & \mbox{\textbf{23.64}{\scriptsize$\pm$1.76}} & \mbox{24.00{\scriptsize$\pm$1.43}} & \mbox{21.69{\scriptsize$\pm$1.87}} & \mbox{\textbf{25.63}{\scriptsize$\pm$1.49}} & \mbox{23.16{\scriptsize$\pm$1.64}} \\
Gallbladder & \mbox{18.78{\scriptsize$\pm$2.21}} & \mbox{16.33{\scriptsize$\pm$2.99}} & \mbox{14.93{\scriptsize$\pm$1.82}} & \mbox{\textbf{39.69}{\scriptsize$\pm$4.57}} & \mbox{24.20{\scriptsize$\pm$3.46}} & \mbox{23.31{\scriptsize$\pm$3.27}} & \mbox{32.85{\scriptsize$\pm$3.68}} & \mbox{\textbf{34.94}{\scriptsize$\pm$3.76}} \\
Esophagus & \mbox{16.29{\scriptsize$\pm$1.02}} & \mbox{18.59{\scriptsize$\pm$2.98}} & \mbox{15.14{\scriptsize$\pm$1.80}} & \mbox{\textbf{19.04}{\scriptsize$\pm$2.68}} & \mbox{48.05{\scriptsize$\pm$1.82}} & \mbox{39.28{\scriptsize$\pm$2.26}} & \mbox{44.92{\scriptsize$\pm$1.79}} & \mbox{\textbf{52.78}{\scriptsize$\pm$1.61}} \\
Stomach & \mbox{4.96{\scriptsize$\pm$0.76}} & \mbox{3.71{\scriptsize$\pm$0.76}} & \mbox{34.63{\scriptsize$\pm$3.53}} & \mbox{\textbf{62.15}{\scriptsize$\pm$3.87}} & \mbox{40.66{\scriptsize$\pm$2.03}} & \mbox{46.67{\scriptsize$\pm$2.05}} & \mbox{52.08{\scriptsize$\pm$1.30}} & \mbox{\textbf{59.89}{\scriptsize$\pm$2.30}} \\
Duodenum & \mbox{3.37{\scriptsize$\pm$1.73}} & \mbox{10.44{\scriptsize$\pm$2.00}} & \mbox{15.04{\scriptsize$\pm$1.46}} & \mbox{\textbf{30.48}{\scriptsize$\pm$2.71}} & \mbox{14.82{\scriptsize$\pm$1.64}} & \mbox{16.41{\scriptsize$\pm$1.62}} & \mbox{\textbf{40.11}{\scriptsize$\pm$1.75}} & \mbox{39.23{\scriptsize$\pm$1.80}} \\
Major vessels & \mbox{24.14{\scriptsize$\pm$1.82}} & \mbox{49.19{\scriptsize$\pm$2.05}} & \mbox{41.22{\scriptsize$\pm$1.65}} & \mbox{\textbf{52.53}{\scriptsize$\pm$1.37}} & \mbox{64.00{\scriptsize$\pm$1.28}} & \mbox{68.46{\scriptsize$\pm$1.18}} & \mbox{68.52{\scriptsize$\pm$0.91}} & \mbox{\textbf{69.56}{\scriptsize$\pm$0.93}} \\
\bottomrule
\end{tabularx}
\end{table}

\paragraph{Cross-dataset 3D segmentation.}
We apply the models trained on MRISegmenter directly to AMOS-MM MRI, without training or tuning on the target dataset. Evaluation covers the ten shared anatomical labels. \methodname achieves the highest transfer Dice (Table~\ref{tab:seg3d_transfer}).

\begin{table}[!htbp]
\centering
\caption{\textbf{Cross-dataset 3D segmentation Dice (\%).} Models trained on MRISegmenter are evaluated directly on AMOS-MM MRI. Values are means $\pm$ 95\% confidence intervals.}
\label{tab:seg3d_transfer}
\footnotesize
\setlength{\tabcolsep}{3pt}
\begin{tabularx}{\linewidth}{@{}l*{2}{>{\centering\arraybackslash}X}@{}}
\toprule
Model & MRISegmenter & AMOS-MM MRI \\
\midrule
PUMIT & 44.57$\pm$0.76 & 31.44$\pm$1.52 \\
MedVAE-3D & 43.70$\pm$0.80 & 33.52$\pm$3.82 \\
AToken & 54.63$\pm$0.68 & 44.39$\pm$1.75 \\
\methodname{} & \textbf{57.20}$\pm$0.76 & \textbf{47.24}$\pm$1.47 \\
\bottomrule
\end{tabularx}
\end{table}

\paragraph{Image synthesis.}
We evaluate how the quality of a tokenizer's latent space influences text-conditioned medical image synthesis. Each frozen tokenizer is integrated into a latent diffusion framework~\cite{rombach2022high} as the only variable: a UNet denoising backbone following the Stable Diffusion v1.4 architecture is randomly initialized for each tokenizer (with input/output channels adapted to match the latent dimensionality), and a frozen CLIP-based text encoder provides conditioning. We train separate diffusion models per tokenizer using identical data, architecture, and hyperparameters, ensuring that differences in generation quality are attributable solely to the tokenizer.

Training uses 70,000 images from GMAI-VL-5.5M~\cite{li2026gmai} spanning seven imaging modalities (10,000 images each). The noise scheduler is DDPM with 1,000 timesteps, a scaled-linear beta schedule, and $\epsilon$-prediction parameterization. The model is trained with an effective batch size of 256, the AdamW optimizer, a constant learning rate with 5\% linear warmup, and classifier-free guidance dropout of 10\%.
Note that the purpose of this experiment is to \emph{compare the generative suitability of different tokenizer latent spaces}, rather than to build state-of-the-art text-to-image generative models or to benchmark fine-grained synthesis. The original paired descriptions in GMAI-VL-5.5M are often long and fine-grained; using them directly may introduce confounding factors from text complexity. We therefore simplify the setup, replacing each caption with a controlled template, \eg, “A medical \{modality\} image,” and keep this protocol identical for all tokenizers. Under this setting, differences in generation quality more directly reflect how well each frozen tokenizer organizes medical images into a latent space that is compact, learnable, and amenable to latent diffusion.

At inference, we use DDIM sampling with 100 denoising steps. We generate 500 images per modality and report the Kernel Inception Distance (KID)~\citep{binkowski2018kid}, Fréchet Inception Distance (FID)~\citep{heusel2017fid}, and Inception Score (IS)~\citep{salimans2016inceptionscore} for evaluation in Table~\ref{tab:syn_full}. \methodname consistently achieves the best average performance across these metrics.

\begin{table*}[t]
\centering
\caption{\textbf{Additional results on image synthesis evaluation.}}
\label{tab:syn_full}
\footnotesize
\begin{tabularx}{\textwidth}{ll*{8}{>{\centering\arraybackslash}X}}
\toprule
Metrics & Method & CFP & CT & Derm. & Endo. & MRI & Path. & X-ray & Avg. \\
\midrule
\multirow{4}{*}{KID ($\downarrow$)} & PUMIT~\citep{luo2023pumit} & \underline{0.098} & 0.110 & 0.096 & 0.110 & 0.067 & 0.296 & 0.080 & 0.123 \\
 & MedVAE~\citep{varma2025medvae} & 0.116 & \underline{0.084} & 0.135 & 0.154 & 0.077 & 0.288 & 0.068 & 0.132 \\
 & MedITok~\citep{ma2025meditok} & 0.256 & 0.099 & \underline{0.078} & \textbf{0.024} & \underline{0.054} & \textbf{0.055} & \underline{0.055} & \underline{0.089} \\
 & \cellcolor{gray!20}\methodname{} & \cellcolor{gray!20}\textbf{0.027} & \cellcolor{gray!20}\textbf{0.076} & \cellcolor{gray!20}\textbf{0.077} & \cellcolor{gray!20}\underline{0.035} & \cellcolor{gray!20}\textbf{0.048} & \cellcolor{gray!20}\underline{0.139} & \cellcolor{gray!20}\textbf{0.019} & \cellcolor{gray!20}\textbf{0.060} \\
\midrule
\multirow{4}{*}{FID ($\downarrow$)} & PUMIT~\citep{luo2023pumit} & 140.40 & 183.49 & 272.26 & 191.49 & 227.64 & 377.12 & 200.14 & 227.51 \\
 & MedVAE~\citep{varma2025medvae} & 118.13 & 154.42 & 202.09 & 170.29 & 145.25 & 325.87 & 88.55 & 172.08 \\
 & MedITok~\citep{ma2025meditok} & \underline{90.65} & \textbf{142.46} & \underline{176.94} & \textbf{58.38} & \underline{122.96} & \underline{99.66} & \underline{77.24} & \underline{109.76} \\
 & \cellcolor{gray!20}\methodname{} & \cellcolor{gray!20}\textbf{59.85} & \cellcolor{gray!20}\underline{146.39} & \cellcolor{gray!20}\textbf{142.87} & \cellcolor{gray!20}\underline{82.39} & \cellcolor{gray!20}\textbf{116.78} & \cellcolor{gray!20}\textbf{94.65} & \cellcolor{gray!20}\textbf{69.17} & \cellcolor{gray!20}\textbf{101.73} \\
\midrule
\multirow{4}{*}{IS ($\uparrow$)} & PUMIT~\citep{luo2023pumit} & 1.024 & 1.031 & 1.077 & \underline{1.048} & 1.049 & 1.066 & 1.057 & 1.050 \\
 & MedVAE~\citep{varma2025medvae} & 1.035 & \underline{1.060} & 1.069 & 1.035 & 1.071 & 1.061 & 1.073 & 1.058 \\
 & MedITok~\citep{ma2025meditok} & \underline{1.039} & 1.048 & \underline{1.104} & 1.038 & \textbf{1.079} & \underline{1.154} & \textbf{1.076} & \underline{1.077} \\
 & \cellcolor{gray!20}\methodname{} & \cellcolor{gray!20}\textbf{1.044} & \cellcolor{gray!20}\textbf{1.060} & \cellcolor{gray!20}\textbf{1.127} & \cellcolor{gray!20}\textbf{1.054} & \cellcolor{gray!20}\underline{1.075} & \cellcolor{gray!20}\textbf{1.159} & \cellcolor{gray!20}\underline{1.074} & \cellcolor{gray!20}\textbf{1.085} \\
\bottomrule
\end{tabularx}
\end{table*}

\paragraph{Visual question answering.}
We evaluate each frozen tokenizer as a visual backbone within the LLaVA-Med~\cite{li2023llavamed} framework. Given a medical image, the tokenizer encodes it into patch-level tokens, which are then fed through a two-layer MLP connector into Mistral-7B-Instruct as the language model backbone. We train separate LLaVA-Med variants using identical data and hyperparameters while keeping the tokenizer frozen.

Training follows a two-stage protocol. In the first stage, only the MLP connector is trained on approximately 509,369 image--text pairs (learning rate $10^{-3}$, one epoch), with 500,000 pairs from PubMedVision~\citep{chen2024huatuogpt} and 9,369 pairs from MM-Skin~\citep{zeng2025mm}. In the second stage, LoRA adapters (rank$=$128, $\alpha=$256, dropout$=$0.05) are applied to all linear layers of the language model and fine-tuned jointly with the connector (learning rate $10^{-4}$ for LoRA, $2 {\times} 10^{-5}$ for the connector, one epoch), using 140,151 VQA items for training (70,000 from PubMedVision; 21,778 from the MM-Skin training split; 28,719 from the Kvasir-VQA-x1 training split; 19,654 from the PathVQA training split).

At inference, we use sampling with a temperature of $0.2$. Model outputs are evaluated via LLM-as-judge binary grading using Qwen3-235B-A22B~\citep{yang2025qwen3}. We train each LLaVA-Med variant with three random seeds and report accuracy as mean $\pm$ 95\% confidence interval in Table~\ref{tab:vqa}.

\begin{figure}[t]
    \centering
    \includegraphics[width=\linewidth]{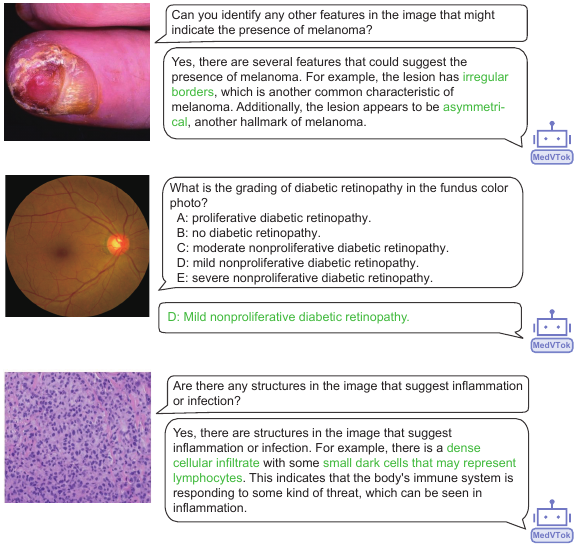}
    \caption{Representative responses from the LLaVA-Med model integrating \methodname as the visual backbone.}
    \label{fig:vqa_cases}
\end{figure}

Two representative cases of LLaVA-Med output (integrating \methodname as the visual encoder) are presented in Fig.~\ref{fig:vqa_cases}, where the model identifies the pathological features in the given medical image and answers the question correctly.

\paragraph{VQA controls.}
We hold the connector, language model, data, LoRA settings, decoding, and judge fixed across tokenizers. We also evaluate a frozen language model with only the connector trained, and re-score the standard model outputs with GPT-5. \methodname leads under connector-only training, and replacing the Qwen3-235B-A22B judge with GPT-5 preserves the tokenizer ranking (Table~\ref{tab:vqa_controls}).

\begin{table}[!htbp]
\centering
\caption{\textbf{VQA accuracy (\%) under training and judge controls.} Connector-only training freezes the language model and uses Qwen3-235B-A22B for grading. Standard training updates the connector and language-model LoRA adapters. Values are means $\pm$ 95\% confidence intervals.}
\label{tab:vqa_controls}
\footnotesize
\setlength{\tabcolsep}{3pt}
\begin{tabularx}{\linewidth}{@{}l*{3}{>{\centering\arraybackslash}X}@{}}
\toprule
Model & \shortstack{Connector only\\Qwen3 judge} & \shortstack{Standard\\Qwen3 judge} & \shortstack{Standard\\GPT-5 judge} \\
\midrule
PUMIT & 23.05$\pm$0.44 & 41.94$\pm$0.33 & 41.83$\pm$0.33 \\
MedVAE & 17.88$\pm$0.48 & 39.93$\pm$0.41 & 39.58$\pm$0.42 \\
MedITok & 24.76$\pm$0.39 & 45.39$\pm$0.28 & 44.82$\pm$0.26 \\
\methodname{} & \textbf{29.86}$\pm$0.36 & \textbf{46.48}$\pm$0.22 & \textbf{46.02}$\pm$0.24 \\
\bottomrule
\end{tabularx}
\end{table}

\begin{table*}[t]
\centering
\caption{\textbf{Additional results on medical report generation evaluation.}}
\label{tab:mrg_full}
\footnotesize
\begin{tabularx}{\textwidth}{ll*{4}{>{\centering\arraybackslash}X}}
\toprule
Metrics & Method & Abdomen & Chest & Pelvis & Avg. \\
\midrule
\multirow{4}{*}{BLEU ($\uparrow$)}
 & PUMIT~\citep{luo2023pumit} & \textbf{0.289}{\scriptsize$\pm$0.008} & 0.193{\scriptsize$\pm$0.003} & 0.164{\scriptsize$\pm$0.007} & 0.215{\scriptsize$\pm$0.005} \\
 & MedVAE-3D~\citep{varma2025medvae} & 0.249{\scriptsize$\pm$0.038} & 0.185{\scriptsize$\pm$0.020} & 0.152{\scriptsize$\pm$0.010} & 0.195{\scriptsize$\pm$0.013} \\
 & MedITok~\citep{ma2025meditok} & 0.256{\scriptsize$\pm$0.013} & \underline{0.219}{\scriptsize$\pm$0.020} & \textbf{0.294}{\scriptsize$\pm$0.017} & \underline{0.257}{\scriptsize$\pm$0.012} \\
 & \cellcolor{gray!20}\methodname{} & \cellcolor{gray!20}\underline{0.272}{\scriptsize$\pm$0.005} & \cellcolor{gray!20}\textbf{0.263}{\scriptsize$\pm$0.045} & \cellcolor{gray!20}\underline{0.288}{\scriptsize$\pm$0.017} & \cellcolor{gray!20}\textbf{0.274}{\scriptsize$\pm$0.009} \\
\midrule
\multirow{4}{*}{ROUGE-L ($\uparrow$)}
 & PUMIT~\citep{luo2023pumit} & \textbf{0.415}{\scriptsize$\pm$0.005} & 0.277{\scriptsize$\pm$0.005} & 0.207{\scriptsize$\pm$0.007} & 0.300{\scriptsize$\pm$0.003} \\
 & MedVAE-3D~\citep{varma2025medvae} & 0.398{\scriptsize$\pm$0.004} & 0.266{\scriptsize$\pm$0.003} & 0.183{\scriptsize$\pm$0.004} & 0.282{\scriptsize$\pm$0.000} \\
 & MedITok~\citep{ma2025meditok} & 0.389{\scriptsize$\pm$0.007} & \underline{0.299}{\scriptsize$\pm$0.025} & \underline{0.407}{\scriptsize$\pm$0.033} & \underline{0.365}{\scriptsize$\pm$0.016} \\
 & \cellcolor{gray!20}\methodname{} & \cellcolor{gray!20}\underline{0.410}{\scriptsize$\pm$0.006} & \cellcolor{gray!20}\textbf{0.353}{\scriptsize$\pm$0.023} & \cellcolor{gray!20}\textbf{0.442}{\scriptsize$\pm$0.011} & \cellcolor{gray!20}\textbf{0.402}{\scriptsize$\pm$0.005} \\
\midrule
\multirow{4}{*}{METEOR ($\uparrow$)}
 & PUMIT~\citep{luo2023pumit} & \textbf{0.353}{\scriptsize$\pm$0.005} & 0.213{\scriptsize$\pm$0.003} & 0.189{\scriptsize$\pm$0.007} & 0.252{\scriptsize$\pm$0.004} \\
 & MedVAE-3D~\citep{varma2025medvae} & 0.318{\scriptsize$\pm$0.025} & 0.221{\scriptsize$\pm$0.010} & 0.172{\scriptsize$\pm$0.008} & 0.237{\scriptsize$\pm$0.014} \\
 & MedITok~\citep{ma2025meditok} & 0.318{\scriptsize$\pm$0.014} & \underline{0.279}{\scriptsize$\pm$0.004} & \underline{0.388}{\scriptsize$\pm$0.021} & \underline{0.328}{\scriptsize$\pm$0.014} \\
 & \cellcolor{gray!20}\methodname{} & \cellcolor{gray!20}\underline{0.336}{\scriptsize$\pm$0.005} & \cellcolor{gray!20}\textbf{0.327}{\scriptsize$\pm$0.045} & \cellcolor{gray!20}\textbf{0.399}{\scriptsize$\pm$0.019} & \cellcolor{gray!20}\textbf{0.354}{\scriptsize$\pm$0.007} \\
\midrule
\multirow{4}{*}{RaTEScore ($\uparrow$)}
& PUMIT~\citep{luo2023pumit} & 0.604{\scriptsize$\pm$0.005} & \underline{0.511}{\scriptsize$\pm$0.035} & 0.201{\scriptsize$\pm$0.015} & 0.439{\scriptsize$\pm$0.006} \\
& MedVAE-3D~\citep{varma2025medvae} & \underline{0.613}{\scriptsize$\pm$0.022} & 0.497{\scriptsize$\pm$0.053} & 0.187{\scriptsize$\pm$0.003} & 0.432{\scriptsize$\pm$0.011} \\
 & MedITok~\citep{ma2025meditok} & 0.610{\scriptsize$\pm$0.006} & 0.501{\scriptsize$\pm$0.008} & \underline{0.594}{\scriptsize$\pm$0.031} & \underline{0.568}{\scriptsize$\pm$0.015} \\
 & \cellcolor{gray!20}\methodname{} & \cellcolor{gray!20}\textbf{0.619}{\scriptsize$\pm$0.005} & \cellcolor{gray!20}\textbf{0.570}{\scriptsize$\pm$0.074} & \cellcolor{gray!20}\textbf{0.612}{\scriptsize$\pm$0.006} & \cellcolor{gray!20}\textbf{0.600}{\scriptsize$\pm$0.006} \\
\midrule
\multirow{4}{*}{GREEN ($\uparrow$)}
 & PUMIT~\citep{luo2023pumit} & \underline{0.421}{\scriptsize$\pm$0.017} & 0.204{\scriptsize$\pm$0.010} & 0.124{\scriptsize$\pm$0.019} & 0.250{\scriptsize$\pm$0.009} \\
 & MedVAE-3D~\citep{varma2025medvae} & 0.401{\scriptsize$\pm$0.024} & 0.186{\scriptsize$\pm$0.003} & 0.105{\scriptsize$\pm$0.064} & 0.230{\scriptsize$\pm$0.179} \\
 & MedITok~\citep{ma2025meditok} & 0.412{\scriptsize$\pm$0.025} & \underline{0.256}{\scriptsize$\pm$0.010} & \underline{0.492}{\scriptsize$\pm$0.026} & \underline{0.386}{\scriptsize$\pm$0.013} \\
 & \cellcolor{gray!20}\methodname{} & \cellcolor{gray!20}\textbf{0.425}{\scriptsize$\pm$0.005} & \cellcolor{gray!20}\textbf{0.300}{\scriptsize$\pm$0.078} & \cellcolor{gray!20}\textbf{0.515}{\scriptsize$\pm$0.044} & \cellcolor{gray!20}\textbf{0.413}{\scriptsize$\pm$0.020} \\
\bottomrule
\end{tabularx}
\end{table*}

\paragraph{Medical report generation.}
We use the AMOS-MM~\cite{ji2022amos} dataset, which provides CT volumes paired with radiology reports covering three anatomical regions (abdomen, chest, and pelvis), with 1,287 pairs for training and 400 pairs for evaluation. CT intensities are windowed to $[-160, 240]$~HU.
Each frozen visual tokenizer is integrated into the LaMed framework~\cite{bai2024m3d} as the visual backbone: given a 3D volume, the tokenizer first encodes it into latent tokens, which are then fed through a connector (a spatial pooling layer that applies 3D average pooling with $2{\times}$ downsampling per dimension) and into Llama-3.1-8B-Instruct adapted with LoRA (rank$=$8, $\alpha=$16, dropout$=$0.1).

We train separate LaMed variants, each with a different visual tokenizer as the frozen backbone, using identical data and hyperparameters, and evaluate all variants on the same test set under the same inference settings.
Training runs for 15 epochs with AdamW (learning rate $5 {\times} 10^{-5}$, cosine decay, 3\% warmup), an effective batch size of 4, and BFloat16 precision. 

At inference, the model generates region-specific findings autoregressively with a temperature of $T=0.2$ and top-$p=0.9$.
We also include MedITok~\cite{ma2025meditok}, a 2D-only tokenizer, for comparison. To adapt it to 3D inputs, we treat the depth dimension as a batch of 2D slices, encode them independently in parallel with the MedITok encoder, and concatenate the resulting tokens along the depth axis before passing them to the connector and LLM. 

\begin{figure}[t]
    \centering
    \includegraphics[width=\linewidth]{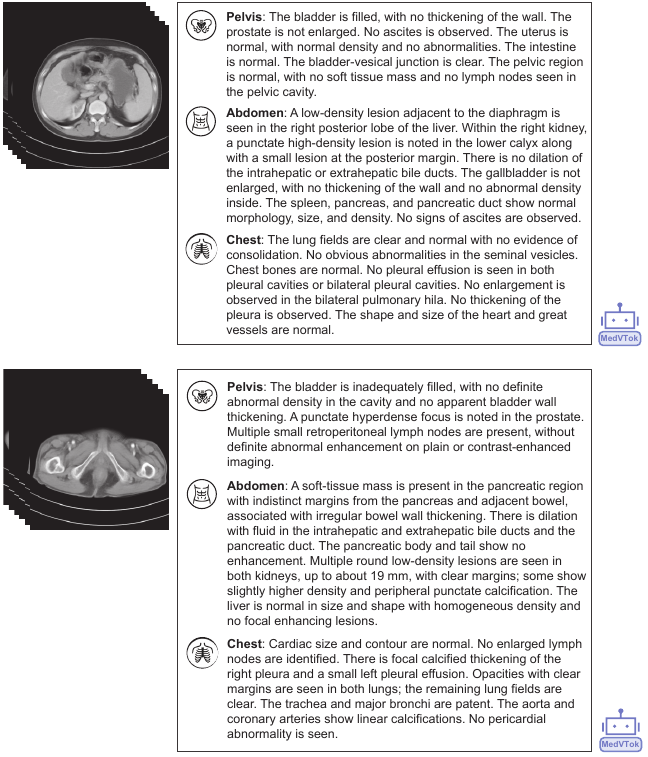}
    \caption{Representative medical reports generated by the LaMed model integrating \methodname as the visual backbone.}
    \label{fig:mrg_cases}
\end{figure}

In Tables~\ref{tab:mrg} and~\ref{tab:mrg_full}, we report BLEU-2, ROUGE-L, METEOR, RaTEScore, and GREEN per anatomical region and globally, as mean $\pm$ 95\% confidence interval over three random seeds. The LaMed model integrating \methodname as the visual encoder achieves superior performance in terms of the average score on each metric. Representative outputs are shown in Fig.~\ref{fig:mrg_cases}, where \methodname directly encodes the whole CT volume and documents the lesions and abnormalities present.

\begin{figure}[t]
    \centering
    \includegraphics[width=\linewidth]{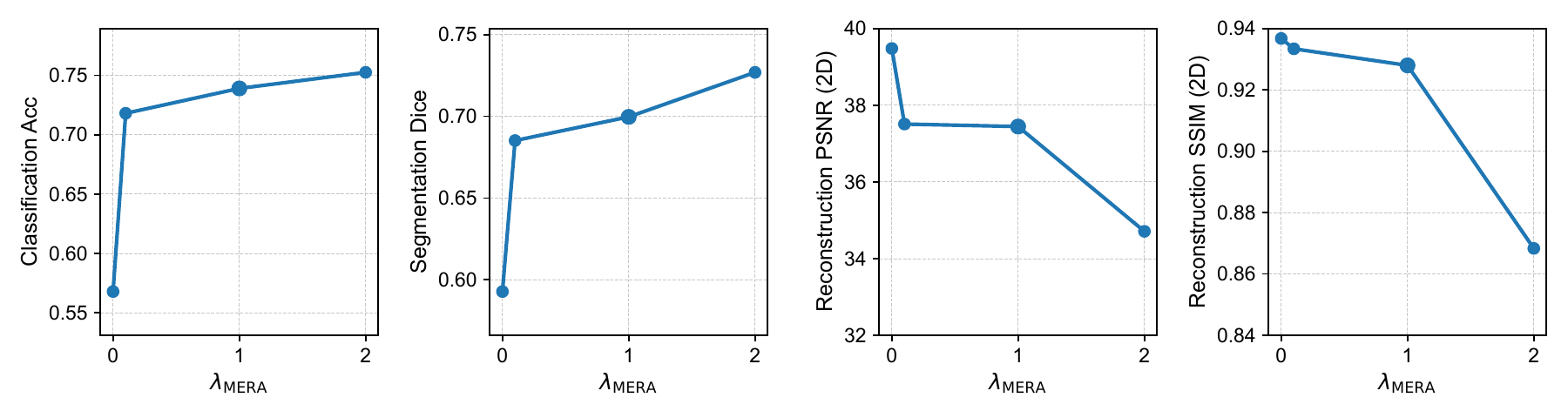}
    \caption{\textbf{Ablation study of $\lambda_\text{MERA}$.} Classification accuracy and segmentation Dice improve with increasing MERA weight, while reconstruction fidelity (PSNR, SSIM) gradually degrades. $\lambda_\text{MERA}=1$ balances semantic quality and reconstruction.}
    \label{fig:ablation_lambda_mera}
\end{figure}

\begin{figure}[!ht]
    \centering
    \includegraphics[width=\linewidth]{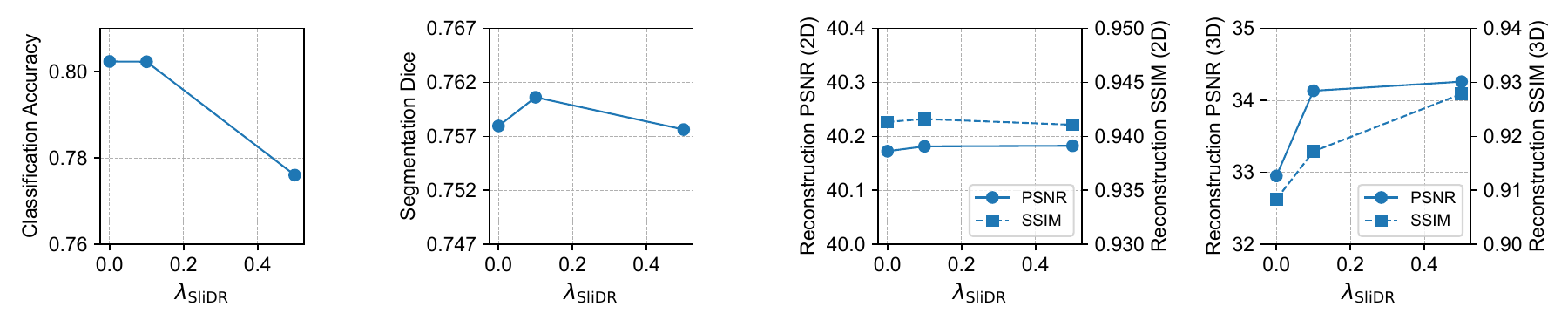}
    \caption{\textbf{Ablation study of $\lambda_\text{SliDR}$.} SliDR improves 3D reconstruction without affecting 2D fidelity. $\lambda_\text{SliDR}=0.1$ provides the best trade-off, as larger values degrade classification (left).}
    \label{fig:ablation_lambda_stc}
\end{figure}

\subsection{Additional Analyses}\label{sec:app:abl}

\paragraph{Effect of $\lambda_\text{MERA}$.}
Fig.~\ref{fig:ablation_lambda_mera} shows the effect of the MERA loss weight for values in $\{0, 0.1, 1, 2\}$. Without semantic alignment ($\lambda_\text{MERA}=0$), classification and segmentation scores remain significantly low despite strong reconstruction, as the tokenizer only learns the low-level structure without discriminating the high-level semantics. Increasing $\lambda_\text{MERA}$ monotonically improves semantic metrics, but at the cost of reconstruction fidelity: at $\lambda_\text{MERA}=2$, PSNR drops by nearly 3~dB and SSIM falls to 0.868. We set $\lambda_\text{MERA}=1$ as the default, which achieves strong semantic performance while maintaining reconstruction quality.

\paragraph{Effect of $\lambda_\text{SliDR}$.}
Fig.~\ref{fig:ablation_lambda_stc} shows the effect of the SliDR loss weight. As expected, 2D reconstruction is unaffected by $\lambda_\text{SliDR}$ since SliDR operates only on volumetric data. For 3D reconstruction, both PSNR and SSIM improve substantially when SliDR is enabled. Further increasing the weight to $\lambda_\text{SliDR}=0.5$ yields marginal 3D gains but degrades classification accuracy by 2.6\%, indicating that excessive volumetric regularization interferes with semantic learning. We therefore adopt $\lambda_\text{SliDR}=0.1$, which captures most of the 3D benefit without compromising other objectives.

\paragraph{Comparison with alternative depth regularizers.}
We compare SliDR with direct gradient matching, total variation (TV), and Laplacian regularization. All variants are initialized from the same 2D-trained \methodname checkpoint and trained on a controlled subset of 10K volumes. We evaluate volumetric reconstruction on AMOS-MM CT and MRISegmenter MRI (Table~\ref{tab:slidr_variants}). SliDR achieves the highest mean PSNR and SSIM on both datasets under this setting.

TV and Laplacian regularization penalize first- and second-order depth differences of the reconstructed volume, respectively, without reference to the ground truth. These penalties encourage smoothness even where the input contains genuine inter-slice changes. SliDR instead matches both orders of slice differentials to the ground-truth volume (Eq.~\ref{eq:stc}), providing a reconstruction target for anatomical transitions.

\begin{table}[!thbp]
\centering
\caption{\textbf{Comparison of depth regularization objectives.} All variants use the same 2D-trained checkpoint and a controlled 10K-volume training subset. Higher PSNR and SSIM indicate better reconstruction. The best mean in each column is marked in \textbf{bold}.}
\label{tab:slidr_variants}
\small
\setlength{\tabcolsep}{5pt}
\begin{tabularx}{\linewidth}{@{}l*{4}{>{\centering\arraybackslash}X}@{}}
\toprule
& \multicolumn{2}{c}{AMOS-MM CT} & \multicolumn{2}{c}{MRISegmenter MRI} \\
\cmidrule(lr){2-3}\cmidrule(lr){4-5}
Objective & PSNR (dB) & SSIM (\%) & PSNR (dB) & SSIM (\%) \\
\midrule
Direct gradient & 33.24{\scriptsize$\pm$0.22} & 93.01{\scriptsize$\pm$0.25} & 26.11{\scriptsize$\pm$0.14} & 78.62{\scriptsize$\pm$0.44} \\
TV & 33.05{\scriptsize$\pm$0.21} & 93.08{\scriptsize$\pm$0.24} & 26.04{\scriptsize$\pm$0.14} & 78.00{\scriptsize$\pm$0.45} \\
Laplacian & 32.99{\scriptsize$\pm$0.21} & 93.08{\scriptsize$\pm$0.24} & 26.20{\scriptsize$\pm$0.15} & 77.86{\scriptsize$\pm$0.45} \\
SliDR & \textbf{33.46}{\scriptsize$\pm$0.22} & \textbf{93.21}{\scriptsize$\pm$0.25} & \textbf{26.23}{\scriptsize$\pm$0.15} & \textbf{78.70}{\scriptsize$\pm$0.44} \\
\bottomrule
\end{tabularx}
\end{table}

\begin{table}[!thbp]
\centering
\caption{\textbf{Classification accuracy (\%) on the unseen PBC modality.} Both protocols use frozen tokenizer features. Values are means $\pm$ 95\% confidence intervals.}
\label{tab:pbc_transfer}
\footnotesize
\setlength{\tabcolsep}{3pt}
\begin{tabularx}{\linewidth}{@{}l*{2}{>{\centering\arraybackslash}X}@{}}
\toprule
Model & $k$-NN & Linear probe \\
\midrule
MedVAE & 40.47$\pm$0.15 & 79.47$\pm$1.35 \\
PUMIT & 36.98$\pm$0.55 & 83.25$\pm$0.84 \\
MedITok & 90.73$\pm$0.62 & 94.62$\pm$0.76 \\
AToken & 92.00$\pm$2.07 & 92.84$\pm$0.87 \\
\methodname{} & \textbf{94.78}$\pm$1.81 & \textbf{96.84}$\pm$0.59 \\
\bottomrule
\end{tabularx}
\end{table}

\begin{table}[!thbp]
\centering
\caption{\textbf{Comparison on the same 2M training images.}}
\label{tab:matched_data}
\footnotesize
\setlength{\tabcolsep}{3pt}
\begin{tabularx}{\linewidth}{@{}l*{3}{>{\centering\arraybackslash}X}@{}}
\toprule
Model & PSNR (dB) & Accuracy (\%) & Dice (\%) \\
\midrule
AToken & 39.35 & 63.65 & 69.57 \\
MedITok & 36.83 & 66.59 & 67.01 \\
\methodname{} & \textbf{39.71} & \textbf{71.70} & \textbf{73.21} \\
\bottomrule
\end{tabularx}
\end{table}

\paragraph{Transfer to an unseen modality.}
We evaluate frozen tokenizer features on PBC~\cite{acevedo2020bloodmnist}, a blood-cell microscopy dataset. This dataset and its hematology modality are absent from tokenizer pretraining. We use $k$-NN classification and train a linear classifier on frozen features. \methodname achieves the highest accuracy under both protocols: 94.78\% and 96.84\%, respectively (Table~\ref{tab:pbc_transfer}).

\paragraph{Comparison on the same training data.}
For controlled comparison with state-of-the-art tokenizers, we retrain AToken, MedITok, and \methodname on MedITok's 2M image--text dataset using the same training settings. AToken and MedITok use text supervision; \methodname uses the images alone. \methodname achieves the highest reconstruction PSNR, classification accuracy, and segmentation Dice in this comparison (Table~\ref{tab:matched_data}).

\begin{figure}
    \centering
    \includegraphics[width=\linewidth]{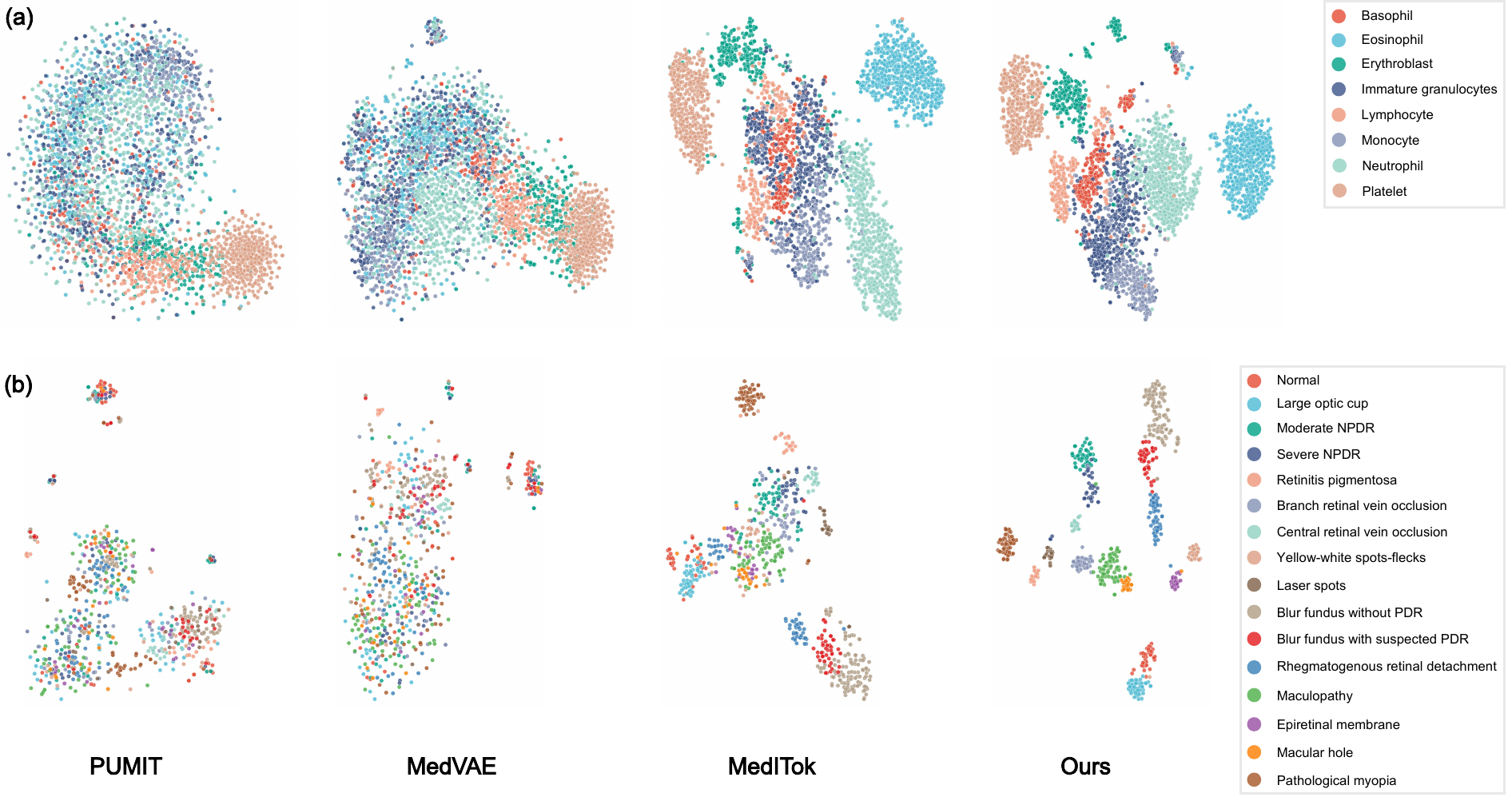}
    \caption{\textbf{$t$-SNE visualization of frozen tokenizer representations.} (a) Out-of-distribution microscopy blood cell dataset. (b) JSIEC fundus dataset (showing the top 16 retinal conditions). \methodname produces the most clearly separated and coherent clusters in both settings.}
    \label{fig:tsne}
\end{figure}

\paragraph{$t$-SNE analysis.}
To qualitatively assess the discriminative structure of the learned representations, we visualize $t$-SNE embeddings of frozen tokenizer features in Fig.~\ref{fig:tsne}. We compare PUMIT, MedVAE, MedITok, and \methodname on two datasets: (a) a microscopy blood cell dataset PBC~\cite{acevedo2020bloodmnist} (covering multiple cell types: basophil, eosinophil, erythroblast, immature granulocytes, lymphocyte, monocyte, neutrophil, platelet), which is an out-of-distribution modality not seen during training, and (b) the JSIEC~\cite{cen2021automatic} fundus dataset. On the PBC dataset, \methodname produces clearly separated clusters for each cell type despite never seeing microscopy images during training, indicating strong generalization of the learned semantic structure. On JSIEC, \methodname forms tighter, more coherent clusters for fine-grained retinal conditions compared to the baselines, which exhibit more inter-class overlap. These visualizations corroborate the quantitative classification and retrieval results, confirming that \methodname encodes semantically meaningful representations that generalize across both in-distribution and out-of-distribution modalities.

\section{Broader Impact}\label{sec:app:broader_impact}
A general-purpose medical visual tokenizer can accelerate multimodal medical AI research and patient-level modeling by providing a shared, reusable visual interface across imaging modalities. This may lower the barrier for building clinical decision-support tools, particularly for under-resourced specialties or institutions that lack the data and compute. By open-sourcing the model and code, we aim to promote reproducibility and equitable access to medical AI infrastructure.

As with any medical AI component, there is a risk that downstream systems built on \methodname could produce incorrect predictions that, if acted upon without appropriate clinical oversight, may lead to patient harm. 
We will release \methodname under the Apache 2.0 license with clear documentation stating that it is intended for research use. We encourage downstream users to rigorously validate performance on their target populations and to comply with applicable regulations before any clinical deployment.

\clearpage

\clearpage

\clearpage

\clearpage

{\footnotesize\setlength{\tabcolsep}{2pt}
\begingroup
\footnotesize
\setlength{\tabcolsep}{3pt}
\begin{longtable}{llr}
\caption{2D training datasets.}
\label{tab:datasets_2d} \\
\toprule
Dataset & Modality & Count \\
\midrule
\endfirsthead
\toprule
Dataset & Modality & Count \\
\midrule
\endhead
\midrule
\multicolumn{3}{r}{\textit{Continued on next page}} \\
\endfoot
\bottomrule
\endlastfoot
RSNA STR Pulmonary Embolism Detection~\cite{s1-rsna-str-pulmonary} & CT & 2,854,455 \\
MELA~\cite{s1-mela-p1} & CT & 1,198,641 \\
RibFrac2020~\cite{s1-ribfrac2020} & CT & 675,992 \\
ATM2022~\cite{s1-atm2022} & CT & 419,489 \\
LIDC-IDRI-ALL-CT~\cite{s1-lidc-idri-all-ct} & CT & 403,462 \\
FLARE22~\cite{s1-flare22} & CT & 236,577 \\
QIN-HEADNECK~\cite{s1-qin-headneck} & CT & 171,846 \\
Pediatric-CT-SEG~\cite{s1-pediatric-ct-seg} & CT & 156,111 \\
AbdomenCT1K~\cite{s1-abdomenct1k} & CT & 145,254 \\
TotalSegmentator v2~\cite{s1-totalsegmentator} & CT & 123,420 \\
ACRIN6668~\cite{s1-acrin6668} & CT & 121,154 \\
COVID-19-NY-SBU~\cite{s1-covid-19-ny-sbu} & CT & 116,764 \\
Totalsegmentator~\cite{s1-totalsegmentator} & CT & 107,256 \\
Parse22~\cite{s1-parse22} & CT & 104,622 \\
CTPelvic1k~\cite{s1-ctpelvic1k} & CT & 96,949 \\
LITS~\cite{s1-lits} & CT & 90,524 \\
AutoPET~\cite{s1-autopet} & CT & 81,135 \\
NLST~\cite{s1-nlst} & CT & 76,000 \\
CAD-PE~\cite{s1-cad-pe} & CT & 66,956 \\
MSD Liver~\cite{s1-msd} & CT & 65,104 \\
CTORG~\cite{s1-ct-org} & CT & 64,904 \\
LNDb~\cite{s1-lndb} & CT & 62,969 \\
HNSCC~\cite{s1-hnscc} & CT & 61,077 \\
ANODE09~\cite{s1-anode09} & CT & 58,647 \\
CTSpine1K Full~\cite{s1-ctspine1k} & CT & 57,145 \\
CPTAC-HNSCC~\cite{s1-cptac-hnscc} & CT & 52,300 \\
Lung-PET-CT-Dx~\cite{s1-lung-pet-ct-dx} & CT & 49,055 \\
ACRIN-NSCLC-FDG-PET~\cite{s1-acrin-nsclc-fdg-pet} & CT & 42,772 \\
Head-Neck Cetuximab~\cite{s1-head-neck-cetuximab} & CT & 42,294 \\
ACRIN-FLT-Breast~\cite{s1-acrin-flt-breast} & CT & 41,063 \\
CT-Covid-19-August2020~\cite{s1-ct-covid-19-august2020} & CT & 39,901 \\
Anti-PD-1 MELANOMA~\cite{s1-anti-pd-1-melanoma} & CT & 39,852 \\
MSD Lung~\cite{s1-msd} & CT & 36,684 \\
OSIC Pul Fib Pro~\cite{s1-osic-pul-fib-pro} & CT & 35,231 \\
SPIE AAPM~\cite{spie-aapm-lung-ct} & CT & 34,776 \\
MIDRC-RICORD-1A~\cite{s1-midrc-ricord} & CT & 33,928 \\
SPIE-AAPM-Lung-CT-Challenge~\cite{spie-aapm-lung-ct} & CT & 33,578 \\
TCGA-LUAD~\cite{s1-tcga-luad} & CT & 33,221 \\
MIDRC-RICORD-1b~\cite{s1-midrc-ricord} & CT & 32,953 \\
Anti-PD-1 Lung~\cite{s1-anti-pd-1-lung} & CT & 32,388 \\
CPTAC-LUAD~\cite{s1-cptac-luad} & CT & 30,882 \\
COVID-19-AR~\cite{Desai2020COVIDRural} & CT & 30,304 \\
TCIA-Covid-19~\cite{Saltz2021SBUCOVID} & CT & 30,304 \\
TCGA-LUSC~\cite{tcga-lusc} & CT & 29,954 \\
SegRap2023~\cite{luo2025segrap2023} & CT & 27,932 \\
VerSe20~\cite{SEKUBOYINA2021102166} & CT & 25,860 \\
C4KC-KiTS~\cite{c4kc-kits} & CT & 25,598 \\
LOLA11~\cite{van_ginneken_2021_4708800} & CT & 24,156 \\
EXACT09~\cite{s1-exact09} & CT & 22,747 \\
KiTS2021~\cite{s1-kits2021} & CT & 22,176 \\
NSCLC-Cetuximab~\cite{s1-nsclc-cetuximab} & CT & 21,911 \\
TCGA-KIRC~\cite{tcga-kirc} & CT & 20,630 \\
CPTAC-CM~\cite{cptac-cm} & CT & 18,513 \\
VESSEL2012~\cite{RUDYANTO20141217} & CT & 18,289 \\
TCGA-UCEC~\cite{tcga-ucec} & CT & 17,924 \\
TCGA-OV~\cite{tcga-ov} & CT & 17,917 \\
OPC-Radiomics~\cite{s1-opc-radiomics} & CT & 17,853 \\
CPTAC-SAR~\cite{cptac-sar} & CT & 17,502 \\
Pancreas-CT~\cite{pancreas-ct} & CT & 14,485 \\
FastPET LD~\cite{s1-fastpet-ld} & CT & 14,047 \\
ACRIN-HNSCC-FDG-PET-CT~\cite{s1-acrin-hnscc-fdg-pet-ct} & CT & 13,989 \\
COVID-19-20~\cite{roth2022rapid} & CT & 12,827 \\
RadImageNet~\cite{s1-radimagenet} & CT, MRI, Ultrasound & 2,639,575 \\
Isic2024~\cite{isic-2024-challenge} & Dermoscopy & 618,536 \\
Derm1M~\cite{yan2025derm1m} & Dermoscopy & 412,886 \\
Isic2020~\cite{s1-isic2020} & Dermoscopy & 44,108 \\
Severance~\cite{han2020assessment} & Dermoscopy & 35,170 \\
BCN20000~\cite{hernandez2024bcn20000} & Dermoscopy & 18,946 \\
Fitzpatrick17k~\cite{groh2021evaluating} & Dermoscopy & 16,577 \\
DERM12345~\cite{yilmaz2024derm12345} & Dermoscopy & 12,345 \\
MILK10k~\cite{philipp2025milk10k} & Dermoscopy & 10,480 \\
SurgeNetYoutube~\cite{de2025scaling} & Endoscopy & 2,061,065 \\
EndoVis2023-Surgtoolloc~\cite{s1-endovis2023-surgtoolloc} & Endoscopy & 745,751 \\
SARAS-MESAD~\cite{bawa2021saras} & Endoscopy & 35,132 \\
EndoVis2023-PitVis~\cite{Speidel2023EndoVis} & Endoscopy & 19,967 \\
In-house Fundus Dataset & CFP & 2,811,937 \\
Airogs~\cite{s1-airogs} & CFP & 101,442 \\
KaggleDR~\cite{diabetic-retinopathy-detection} & CFP & 88,712 \\
Eyepacs~\cite{gulshan2016development} & CFP & 35,126 \\
BIOMEDICA~\cite{lozano2025biomedica} & CFP, Dermoscopy, X-ray & 253,443 \\
HISTAI~\cite{nechaev2025histai} & Histopathology & 4,000,000 \\
PathGen~\cite{sun2024pathgen} & Histopathology & 1,606,060 \\
Bone Marrow Cytomorphology~\cite{s1-bone-marrow-cytomorphology} & Histopathology & 27,783 \\
Lysto~\cite{lysto2024} & Histopathology & 14,920 \\
BraTS2023~\cite{s1-brats2023} & MRI & 973,066 \\
IXI~\cite{s1-ixi} & MRI & 924,426 \\
BrainTumour~\cite{s1-braintumour} & MRI & 246,671 \\
Learn2Reg~\cite{s1-learn2reg} & MRI & 89,572 \\
Vestibular-Schwannoma-SEG~\cite{s1-vestibular-schwannoma-seg} & MRI & 51,574 \\
OASIS2~\cite{s1-oasis2} & MRI & 39,741 \\
ATLAS 2~\cite{s1-atlas-2} & MRI & 36,029 \\
GLIS-RT~\cite{glis-rt} & MRI & 24,979 \\
CrossMoDA2023~\cite{Dorent_2023} & MRI & 21,981 \\
CAS2023~\cite{cas2023} & MRI & 19,200 \\
ASC18~\cite{xiong2021global} & MRI & 13,552 \\
OLIVES~\cite{prabhushankar2022olives} & OCT & 152,346 \\
OCT2017~\cite{kermany2018labeled} & OCT & 84,484 \\
APTOS-2021~\cite{zhang2026aptos} & OCT & 56,972 \\
Retina-OCT-C8~\cite{9740985} & OCT & 24,000 \\
ACOUSLIC~\cite{sappia_2024_12697994} & Ultrasound & 210,929 \\
SYSU-CEUS-FLL-intv1~\cite{SYSUCEUSFLL} & Ultrasound & 43,611 \\
CACTUS~\cite{elmekki2025cactus} & Ultrasound & 37,736 \\
RSNA Mammography~\cite{rsnamammography} & X-ray & 54,710 \\
DDSM~\cite{lee2017curated} & X-ray & 20,800 \\
VinDr-Mammo~\cite{nguyen2023vindr} & X-ray & 19,982 \\
MIMIC-CXR-JPG~\cite{johnson2019mimic} & X-ray & 372,693 \\
ReXGradient-160K~\cite{zhang2025rexgradient} & X-ray & 273,004 \\
CheXpertPlus~\cite{s1-chexpertplus} & X-ray & 223,315 \\
PadChest~\cite{bustos2020padchest} & X-ray & 160,839 \\
NIH Chest X-rays~\cite{s1-nih-chest-x-rays} & X-ray & 112,120 \\
COVIDx CXR-4~\cite{s1-covidx-cxr-4} & X-ray & 84,817 \\
BRAX~\cite{reis2022brax} & X-ray & 40,944 \\
Ranzcr clip~\cite{wang2017chestx} & X-ray & 33,665 \\
RSNA PDC~\cite{RSNA2018Pneumonia} & X-ray & 29,684 \\
VinBigData CXR~\cite{nguyen2022vindr} & X-ray & 9,850 \\
\midrule
\textbf{Total} & & \textbf{28,769,502} \\
\end{longtable}
\endgroup

}

\clearpage

{\footnotesize\setlength{\tabcolsep}{2pt}
\begingroup
\footnotesize
\setlength{\tabcolsep}{3pt}
\begin{longtable}{llr}
\caption{3D training datasets.}
\label{tab:datasets_3d} \\
\toprule
Dataset & Modality & Count \\
\midrule
\endfirsthead
\toprule
Dataset & Modality & Count \\
\midrule
\endhead
\midrule
\multicolumn{3}{r}{\textit{Continued on next page}} \\
\endfoot
\bottomrule
\endlastfoot
CT-RATE~\cite{hamamci2026generalist} & CT & 15,000 \\
RSNA STR Pulmonary Embolism Detection~\cite{s1-rsna-str-pulmonary} & CT & 7,148 \\
FLARE22~\cite{s1-flare22} & CT & 1,874 \\
LIDC-IDRI-ALL-CT~\cite{s1-lidc-idri-all-ct} & CT & 967 \\
AbdomenCT1K~\cite{s1-abdomenct1k} & CT & 882 \\
QIN-HEADNECK~\cite{s1-qin-headneck} & CT & 816 \\
ACRIN6668~\cite{s1-acrin6668} & CT & 764 \\
MELA~\cite{s1-mela-p1} & CT & 688 \\
OPC-Radiomics~\cite{s1-opc-radiomics} & CT & 606 \\
CTPelvic1k~\cite{s1-ctpelvic1k} & CT & 525 \\
HNSCC~\cite{s1-hnscc} & CT & 507 \\
TCGA-KIRC~\cite{tcga-kirc} & CT & 474 \\
NSCLC-Cetuximab~\cite{s1-nsclc-cetuximab} & CT & 465 \\
Head-Neck-PET-CT~\cite{s1-head-neck-pet-ct} & CT & 443 \\
RibFrac2020~\cite{s1-ribfrac2020} & CT & 417 \\
Head-Neck Cetuximab~\cite{s1-head-neck-cetuximab} & CT & 399 \\
Totalsegmentator dataset~\cite{s1-totalsegmentator} & CT & 346 \\
C4KC-KiTS~\cite{c4kc-kits} & CT & 335 \\
Pediatric-CT-SEG~\cite{s1-pediatric-ct-seg} & CT & 296 \\
MSD HepaticVessel~\cite{s1-msd} & CT & 286 \\
MSD Pancreas~\cite{s1-msd} & CT & 280 \\
ACRIN-NSCLC-FDG-PET~\cite{s1-acrin-nsclc-fdg-pet} & CT & 278 \\
ACRIN-FLT-Breast~\cite{s1-acrin-flt-breast} & CT & 238 \\
CTSpine1K Full~\cite{s1-ctspine1k} & CT & 229 \\
TCGA-UCEC~\cite{tcga-ucec} & CT & 205 \\
COVID-19-20~\cite{roth2022rapid} & CT & 199 \\
TCGA-LUAD~\cite{s1-tcga-luad} & CT & 188 \\
CPTAC-HNSCC~\cite{s1-cptac-hnscc} & CT & 186 \\
GLIS-RT~\cite{glis-rt} & CT, MRI & 644 \\
MR-RATE~\cite{Forithmus2026MRRATE} & MRI & 20,000 \\
BraTS2023~\cite{s1-brats2023} & MRI & 8,956 \\
BrainTumour~\cite{s1-braintumour} & MRI & 1,936 \\
LLD-MMRI2023~\cite{LLD-MMRI2023} & MRI & 1,580 \\
Learn2Reg~\cite{s1-learn2reg} & MRI & 828 \\
Vestibular-Schwannoma-SEG~\cite{s1-vestibular-schwannoma-seg} & MRI & 484 \\
M\&Ms-2~\cite{martin2023deep} & MRI & 448 \\
CrossMoDA2023~\cite{Dorent_2023} & MRI & 227 \\
\midrule
\textbf{Total} & & \textbf{70,144} \\
\end{longtable}
\endgroup

}

\end{document}